# Folding with a protein's native shortcut network

Susan Khor
slc.khor@gmail.com
Oct 24, 2017

**Abstract**

A complex network approach to protein folding is proposed. The graph object is the network of shortcut edges present in a native-state protein (SCN0). Although SCN0s are found via an intuitive message passing algorithm (S. Milgram, Psychology Today, May 1967 pp. 61-67), they are meaningful enough that the logarithm form of their contact order (SCN0_lnCO) correlates significantly with protein kinetic rates, regardless of protein size. Further, the clustering coefficient of a SCN0 ($C_{SCN0}$) can be used to combine protein segments iteratively within the Restricted Binary Collision model to form the whole native structure. This simple yet surprisingly effective strategy identified reasonable folding pathways for 12 small single-domain two-state folders, and three non-canonical proteins: ACBP (non-two-state), Top7 (non-cooperative) and DHFR (non-single-domain, > 100 residues). For two-state folders, $C_{SCN0}$ is relatable to folding rates, transition-state placement and stability. The influence of $C_{SCN0}$ on folding extends to non-native structures. Moreover, SCN analysis of non-native structures could suggest three fold success factors for the fast folding Villin headpiece peptide. These results support the view of protein folding as a bottom-up hierarchical process guided from above by native-state topology, and could facilitate future constructive demonstrations of this long held hypothesis for larger proteins.

## 1. Introduction

The problem of knowing how a biological sequence of amino acids finds its functional configuration from an extended state in the crowded environment of a cell has been around for at least half a century [1]. The difficulty of this problem is compounded by the impossibility of an unbiased random search solution for the fast folding rates required [2]. Hence, the existence of preferred or dominant protein folding pathways was conjectured, and the problem of what determines a protein's folding pathway(s) arose [3, 4, 5].

An early and still influential response to this challenge is the native-state topology hypothesis, which proposes that protein folding rates and mechanisms are largely determined by the native structure of proteins [6]. A strong evidence for this hypothesis is the significant negative correlation between the folding rates of small proteins and their relative contact order (RCO) [7, 8]. An implication of this hypothesis is that structurally homologous proteins fold in similar ways, and therefore should have similar folding pathways. However, the discovery of two structurally homologous, nearly symmetric proteins (proteins G and L) that fold differently, but with folding rates that are within one order of magnitude from each other [9], motivated a re-visit of and search for other determinants of protein folding pathways.

In this paper, we use a Restricted Binary Collision (RBC) model to explore the applicability of the clustering coefficient of a protein's native shortcut network ($C_{SCN0}$) to identify plausible protein folding pathways. Our RBC model is inspired by Yang and Sze's mesoscopic approach [10]. RBC is in the tradition of theoretical studies that take a low-resolution approach to protein folding, and rely only on native interactions (Gō models) [11, 12, 13]. The clustering coefficient of a network is a measure of the network's cliquishness. Its relevance to protein kinetics was observed previously [14], but only via correlation analysis [15] and peptide bonds are included in [14]. In contrast, we demonstrate constructively, how the clustering coefficient of a network that excludes edges between adjacent residues can provide clues to preferred protein folding pathways. Our findings support the native-state topology hypothesis even for structurally homologous near symmetric proteins.

## 2. Materials and Method

### 2.1 The native shortcut network

A shortcut network (SCN) is a sub-graph of a Protein Residue Network (PRN). The edges of a SCN comprise shortcut edges found by the Euclidean-distance Directed Search (EDS) algorithm on a PRN. PRN construction, the EDS algorithm, and the concept of a shortcut edge, were introduced in [16].

A PRN is a simple, undirected and connected graph. A PRN node represents a protein residue. A PRN edge connects nodes $u$ and $v$ if and only if $|u - v| > 1$, and their interaction strength, $I_{uv} = \frac{n_{uv} \times 100}{\sqrt{R_u \times R_v}} \geq 5.0$. $n_{uv}$ is the number of distinct pairs $(i, j)$ such that $i$ is an atom of residue $u$, $j$ is an atom of residue $v$, and the Euclidean distance between atoms $i$ and $j$ is at most 7.5 Å. $R_u$ and $R_v$ are extracted from a table of normalization values by residue type (Table 1 in [17]). PRN construction is based on refs. [17] and [18], but both main- and side-chain atoms are considered when evaluating interaction strength. The limitations of main-chain only or side-chain only coarse-grained models of protein folding have been contemplated [19, 20]. Our mixed approach is not unique [21, 22, 23].

What is perhaps a little unusual for a protein folding model is peptide bonds are prohibited as PRN edges. This design choice was inherited from [18], and has so far shown itself to be a fortuitous one. It has allowed investigation of allosteric communication in proteins [24], and provided this research is successful, it potentially equips the PRN model to facilitate an integrated investigation into protein folding and dynamics [25]. We note that peptide bond effects, i.e. distance fluctuations (or rather lack thereof) between residues adjacent on a protein sequence, are commonly excluded when examining native protein dynamics: e.g. COMMA's communication pathways [26], and force constant calculation [23]. The inclusion of peptide bonds in PRNs has a detrimental effect on shortcut edges (Fig. S1).

EDS is a greedy local search algorithm similar to Kleinberg's [27] which formalizes the message-passing protocol used in Milgram's social network experiment [28], but with backtracking to exploit a dynamic repository of distance information created anew with each search. An intuitive description of EDS as a message-passing protocol in a social network follows below. A more formal description of shortcut identification via EDS on a PRN is in Note S1. Fig. 1 illustrates the native shortcuts for a protein.

Under the EDS protocol, a message is transmitted from a source to a target with two lists: a participant list, which records persons who have already participated in the message-relay, and a candidate list, which collects identity and distance information of all persons considered to pass the message so far. The two lists are empty at the start of each transmission. As a message-relay progresses, participants in the relay include themselves on the participant list, and add to the candidate list, distance information of their immediate neighbors on the social network to the target recipient. A message intended for Z is passed from A to B if B is not on the participant list yet, and if B is closest to Z amongst all others on the candidate list. Since participants act on partial knowledge of the network, their decisions may not be optimal in the global picture.

At some point during a message-relay, it is possible that X may need to pass the message to Q, but Q is not directly connected to X. In this case, backtracking (a cascade of return-to-sender actions) is used to get the message from X to Q. Backtracking allows a message to retrace its steps and does not violate the EDS protocol. Since Q is on the candidate list X receives (this is how X knows about Q in the first place if Q is not a direct neighbor of X), Q was considered as a participant earlier in the relay but was rejected at that time. Now, however, Q emerges as the best candidate to transmit the message. If Q were an immediate neighbor of X, then the XQ contact acts as a shortcut in this relay because the presence of XQ helps to avoid backtracking. From this informal description of a shortcut, it is clear that every shortcut is part of some cycle in the network.

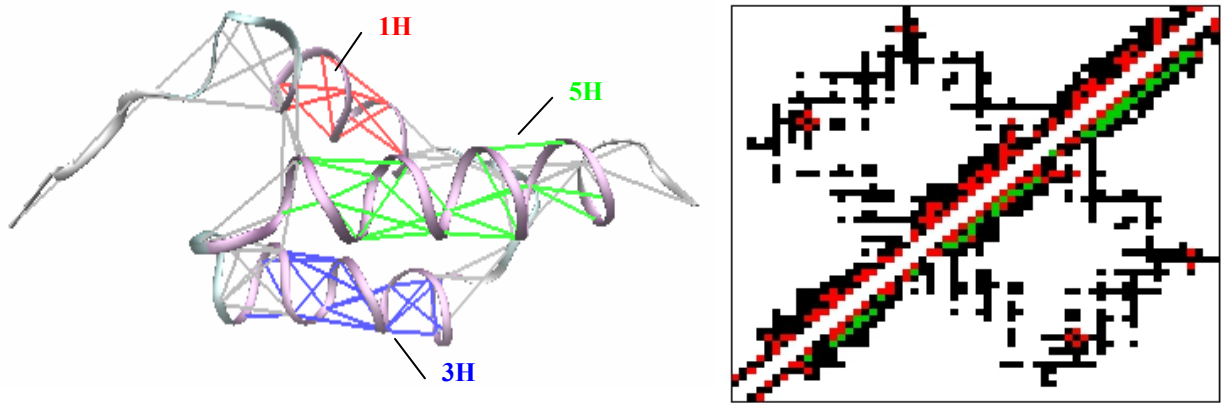

**Fig. 1 Left:** Shortcut edges are drawn on a cartoon of 1BDD with VMD [29]. Shortcuts within the three helix structures 1H, 3H and 5H, are colored red, blue and green respectively. **Right:** Adjacency matrix (contact map) of 1BDD's PRN. Shortcut edges are marked by red cells, non-shortcut edges by black cells, and (in the lower triangle only) hydrogen bonds identified with HBPLUS version 3.2 [30] that coincide with edges by green cells.

## 2.2 SCN clustering coefficient

The clustering coefficient $C$ of a graph $G$, is the average clustering coefficient over all of $G$'s nodes which number $N$: $C_G = \frac{1}{N}\sum_i^N C(i)$. The clustering coefficient of a node $C(i) = \frac{2e_i}{k_i(k_i-1)}$, where $e$ is the number of links between node $i$'s $k$ direct neighbors [31]. $C_{SCN0}$ is clustering coefficient computed with the edges of a native SCN (SCN0) exclusively.

## 2.3 Protein datasets

Proteins and kinetic data in this study are documented in Table S1, Table S2 [32], Table 1 of ref. [33] and Table 1 of ref. [14]. Of the proteins in Table S1, 16 structurally diverse (α, β and α/β) proteins were selected to test the applicability of our RBC with $C_{SCN0}$ method (Table S3). The preferred folding pathways of these proteins are well-described in existing literature, and they highlight different facets of the folding problem.

2IGD and 2PTL are structural homologues, with low (15%) sequence identity, but with different folding pathways [9]. 1SHG and 1SRM are structural homologues, with low (36%) sequence similarity, which have folding transition state ensembles that resemble each other, i.e. similar folding pathways [34]. 2CI2 folds via the nucleation condensation mechanism, wherein secondary and tertiary structures form simultaneously [35], and thus may prove challenging for our bottom-up hierarchical approach. Nevertheless, ref. [36] argued that the nucleation condensation mechanism can be reconciled with the framework model. Young and Sze [10] showed that a coarse-grained framework approach based on energetics can work for 2CI2. 2KJV is a 'complex two-state folder' that exhibits transition state shifts along a pathway, and between pathways [37]. 2KJW is a circular permutant (P54-55) of 2KJV that has the same tertiary structure as 2KJV, but folds via the alternative pathway.

Due to their "special" characteristics (non-two-state, non-cooperative or non-single-domain), we label the proteins discussed in this paragraph *non-canonical*. 2ABD and 1ST7 are respectively, the bovine and yeast structures of the extensively studied four helix bundle ACBP (acyl-coenzyme A-binding protein), whose folding pathway could challenge our infeasible SSU restriction (section 2.5). ACBP is also interesting because, depending on experimental conditions, it can be classified as a multi-state folder [15, 33]. 1QYS is a de novo protein with < 100 residues but exhibits non-cooperative behavior, and has a very stable intermediate structure [38]. Typically, small proteins fold cooperatively, and make one transition from the unfolded state to the native state. 7DFR has two structural domains: a continuous one in the middle of the chain, and a discontinuous one spanning the N- and C- termini. At 159 residues, 7DFR is the largest protein to test our RBC with $C_{SCN0}$ method, hitherto.

Finally, 2F4K is a very short (35 residue) α-helix protein modified for fast folding. It is included to study the influence of native topology, in terms of $C_{SCN0}$, in non-native structures which were generated by Molecular Dynamics (MD) simulations of its folding [41].

### 2.4 Overview of the Restricted Binary Collision (RBC) model

In the RBC model, protein folding begins with an initial set of fully formed, isolated secondary structure elements (SSE), and terminates with the native conformation. In each fold step, exactly two secondary structure units (SSU) are collided to form a new SSU, and this process is iterated until all the SSEs of a protein coalesce into a single SSU. Throughout this assembly process, both SSEs and SSUs are assumed to be rigid units adopting their respective native forms. Such an intuitive bottom-up (local to global) hierarchic process of self-assembly is a long-standing idea in protein folding theory [42, 43].

Our challenge is to identify a sequence of collision steps that can reasonably form a folding pathway for a protein. Reasonableness is evaluated in light of existing protein literature. Since one collision happens at each fold step, the process terminates in $n$-1 steps, where $n$ is the number of SSEs. The number of candidate conformations at each step reduces by one as the search approaches its target (native conformation). The total number of conformations, and thus the number of fitness function evaluations, involved in the search for a folding pathway is $n(n-1)/2$. All the folding pathways in this paper were easily produced on an ordinary laptop PC running un-optimized Visual C++ code on Windows OS.

### 2.5 Secondary structure elements and (in)feasible secondary structure units

A *secondary structure element* (SSE) is a maximal contiguous segment of a protein sequence with the same secondary structure assignment, i.e. either H (helix), S (strand), or T (turn). The SSEs of a protein (Table S3) are arranged in protein sequence order in a ring for RBC (Fig. 2).

A *secondary structure unit* (SSU) comprises one or more non-Turn SSEs, and all Turn SSEs sandwiched between any of its non-Turn SSEs in the ring of SSEs. Table 1 enumerates all possible SSUs for 1BDD with their respective SSE constituents.

A SSU is *infeasible* if it does not respect protein chain continuity. For instance, 1BDD's SSU(1H 5H) is infeasible because its residues do not occupy a contiguous segment of the protein sequence. The purpose of this restriction is to prioritize the closure of smaller over larger loops, and it is aligned with the principal that protein folding prefers paths that reduce the probability of premature conformational entropy loss [44, 45]. However, a similar restriction in other models has been criticized for being un-physical for proteins such as barnase [6, 46].

Infeasible SSUs need not occur just between the terminus ends of proteins. However, this most extreme case of infeasible SSU, early N-C termini coupling, has attracted quite a lot of study, and found to affect folding rates and even the transition-states of some proteins [47, 48, 49]. This restriction can be lifted in RBC; we do so temporarily for ACBP in section 3.4. Allowing N-C termini SSUs modifies almost all of the $C_{SCN0}$ folding pathways reported in section 3.3, and the changes are not all benign (Table S5).

It is important to note that while RBC prohibits formation of infeasible SSUs, it allows interactions between multi-SSE SSUs when calculating Conformation fitness (section 2.7), and so islands of contiguous segments distant from each other on a protein sequence may not be completely isolated from one another in RBC. There is also the reverse problem of a model allowing too much interaction between folded blocks when dealing with multi-domain proteins [40].

From the brief discussion here, it is clear that a number of variations on RBC are possible. This flexibility is common in model design, and it provides room to accommodate different protein folding patterns, if indeed this diversity turns out to be the case. For now, we stick with a simple version to explore the principle of using a structural metric, $C_{SCN0}$, to propose protein folding pathways.

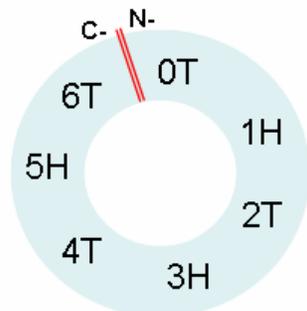

**Fig. 2** SSEs of 1BDD arranged in protein sequence order in a ring to construct SSUs.

**Table 1** Secondary structure units (SSUs) for 1BDD, their constituent SSEs, and associated Conformations. Respectively, #, |SC|, |SCSE| and |SCLE| are the number of residues, shortcuts, short-range shortcuts, and long-range shortcuts in a SSU.

| SSU | SSEs in SSU | # | |SC| | |SCSE| | |SCLE| | Conformation |
|---|---|---|---|---|---|---|
| (1H) | 1H | 8 | 10 | 10 | 0 | (1H) (3H) (5H) |
| (3H) | 3H | 13 | 21 | 21 | 0 | (1H) (3H) (5H) |
| (5H) | 5H | 15 | 20 | 20 | 0 | (1H) (3H) (5H) |
| (1H 3H) | 1H, 2T, 3H | 28 | 41 | 41 | 0 | (1H 3H) (5H) |
| (3H 5H) | 3H, 4T, 5H | 32 | 51 | 51 | 0 | (1H) (3H 5H) |
| (1H 5H) | 0T, 1H, 5H, 6T | 36 | 47 | 43 | 4 | (1H 5H) (3H) |
| (1H 3H 5H) | 0T, 1H, 2T, 3H, 4T, 5H, 6T | 60 | 90 | 84 | 6 | (1H 3H 5H) |

### 2.6 Conformations and their search space

A *Conformation* is a combination of one or more SSUs such that it contains all the non-Turn SSEs of a protein. Table 1 lists all of 1BDD's Conformations. Turn SSEs between SSUs are not added when forming Conformations. For example, 0T, 2T and 6T are excluded from Conformation(1H) (3H 5H), i.e. the residues of these turn segments are ignored when evaluating Conformation fitness.

In general, a (discrete) search space is a graph $G$ whose nodes represent unique combinations and whose edges represent permitted transformations between combinations. Similarly, in a Conformation search space graph, the nodes denote Conformations, and the edges denote (kinetic) *accessibility* of one Conformation from another. A Conformation Y that can be derived from another Conformation X via a binary collision of X's SSUs is accessible from X. For instance Conformation(1H 3H)(5H) is accessible from Conformation(1H)(3H)(5H) because it can be derived by combining (1H) and (3H). Fig. 3 depicts 1BDD's Conformation search space.

Conformations with infeasible SSUs are *infeasible* Conformations, and strictly speaking, are inaccessible in our search space. With the energy function we use (section 2.7), it is common for infeasible Conformations to have relatively high amounts of energy. For instance, 1BDD's Conformation(1H 5H)(3H) which contains the infeasible SSU(1H 5H) has the largest energy (Table 2).

### 2.7 Conformation fitness

A search space becomes a fitness landscape when its nodes are assigned fitness values. Fitness of a Conformation may be defined in several ways. We adapt the linear combination approach used by Yang and Sze [10], and introduce some non-linearity by allowing multi-SSE SSUs (SSUs with more than one non-Turn SSEs) to interact with each other when calculating the fitness of a Conformation. This interaction is for fitness calculation at a single fold step only; it does not produce a new SSU, and interaction between the multi-SSE SSUs may change in future steps. Let X be a Conformation with $p$ single-SSE SSUs, and $q$ multi-SSE SSUs. Then fitness of Confirmation X is:

$$\mathbf{F}(X) = f(RSU = \bigcup_{i=1}^{q} SSU_i) + \sum_{i=1}^{p} f(RS_i = SSU_i) \qquad \text{Eq. 1}$$

where $RSU$ is the union of the residues belonging to the $q$ multi-SSE SSUs in Conformation X, $RS_i$ is the set of residues belonging to $SSE_i$, and $f$ is either the energetic or structural fitness function defined below. Note S2 elaborates on Eq. 1. The exact interactions between multi-SSE SSUs can be refined, as is done for 7DFR (section 3.4).

*Energetic fitness.* To calculate the energy of a set of residues, Yang and Sze [10] extract the PDB coordinates of the residues and apply the original Rosetta energy function to the extracted coordinates. We do the same but use DeepView's (version 4.1.0) Compute Energy (Force Field) function [50], which is a partial implementation of the GROMOS96 force-field, to do the energy calculations. Table 2 lists the energy values of 1BDD's SSUs and Conformations.

*Structural fitness.* The $C_{SCN0}$ value of a set of residues is the clustering coefficient obtained using only the native shortcuts whose endpoints both belong to the set of residues. The $C_{SCN0}$ values for 1BDD's SSUs and Conformations are listed in Table 2.

**Table 2** Energetic and structural fitness of 1BDD's SSUs and Conformations.

| SSU | Conformation | SSEs in Conformation | Energetic fitness (E) | | $C_{SCN0}$ fitness | |
|---|---|---|---|---|---|---|
| | | | SSU | Conformation | SSU | Conformation |
| (1H) | (1H) (3H) (5H) | 1H, 3H, 5H | 25.3863 | 337.5233 | 0.4708 | 1.2089 |
| (3H) | (1H) (3H) (5H) | 1H, 3H, 5H | 157.7012 | 337.5233 | 0.5026 | 1.2089 |
| (5H) | (1H) (3H) (5H) | 1H, 3H, 5H | 154.4357 | 337.5233 | 0.2356 | 1.2089 |
| (1H 3H) | (1H 3H) (5H) | 1H, 2T, 3H, 5H | -244.7240 | -90.2885 | 0.2917 | 0.5272 |
| (3H 5H) | (1H) (3H 5H) | 1H, 3H, 4T, 5H | -222.1320 | -196.7460 | 0.3420 | 0.8128 |
| (1H 5H) | (1H 5H) (3H) | 0T, 1H, 3H, 5H, 6T | 545.1691 | 702.8700 | 0.2620 | 0.7646 |
| (1H 3H 5H) | (1H 3H 5H) | 0T, 1H, 2T, 3H, 4T, 5H, 6T | -271.9150 | -271.9150 | 0.2534 | 0.2534 |

### 2.8 Folding pathway identification

A folding pathway is a path in the Conformation search space graph that starts at the Conformation with the maximum number of SSUs, and ends at the Conformation with only one SSU. The challenge is to identify a fitness gradient or rule to guide the selection of intermediate Conformations that make a plausible folding pathway for proteins in general.

Yang and Sze [10] approached this challenge by following the path of steepest descent in free-energy. The pathway construction rule we propose is: choose an accessible Conformation with the *largest* $C_{SCN0}$ at each step. In Fig. 3, Conformation(1H)(3H 5H) is selected as the intermediate Conformation because it has the largest $C_{SCN0}$ of the feasible Conformations accessible from Conformation(1H)(3H)(5H). For 1BDD, the $C_{SCN0}$ rule also identifies the folding pathway of steepest descent in DeepView's energy values.

## 3. Results and Discussion

### 3.1 Native shortcut edges capture the structural essence of proteins

The purpose of this section is to show that native shortcut networks (SCN0s) are not random contact graphs, even though they are identified via the EDS algorithm which is biophysics ignorant. This is done by examining how several well-established structural characteristics of protein molecules that are relevant to protein folding, are expressed in SCN0s.

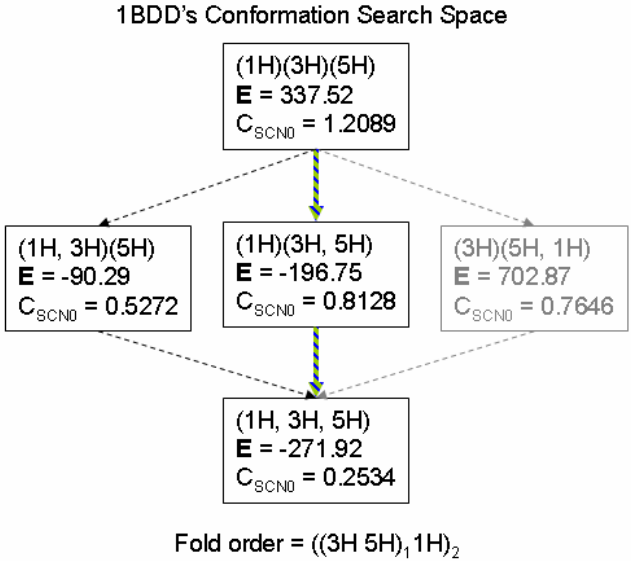

Pathway construction rules for intermediate Conformations:

(i) The **E** rule (blue): choose an accessible Conformation that has the *smallest* energy value.

(ii) The $C_{SCN0}$ rule (green): choose an accessible Conformation with the *largest* $C_{SCN0}$.

**Fig. 3** 1BDD's Conformation search space. The arcs denote accessibility between Conformations. Infeasible and thus inaccessible Conformations are placed in grayed boxes. The $C_{SCN0}$ folding pathway is traced with green arcs, while the **E** (steepest descent in energy) folding pathway is traced with blue arcs. In the case of 1BDD, these two folding pathways are identical, and the overlapping portions of the pathways are shown by the blue and green striped arcs.

A SCN0 covers almost all of the residues in a native-folded protein chain, $N_{SCN0} \approx N$. The number of edges in a SCN0 scales linearly with $N$ (Fig. S1). This linear relationship is not unusual. A similar relationship between protein size and number of contacts is observed with Weikl's contact maps [45].

The breakdown of a SCN0's edges into short-range and long-range reflects a protein's secondary structure makeup (Fig. S3). An edge is short-range if it connects residues that are at most 10 positions apart on a protein sequence, and long-range otherwise. The proportion of native shortcuts which are short-range is larger in α-rich proteins. This is expected since helices are formed with short-range contacts, while medium to long-range contacts are required for beta sheets. The arrangement of native shortcut edges within the ordered secondary structures (α-helix and β-sheet) is more regular than within the disordered regions (loops) (Figs. S4 & S5).

Native shortcut edges are significantly more enriched with hydrogen bonds than any other subset of PRN edges examined (Fig. S6). There is more overlap between shortcut edges and hydrogen bonds within the α-helix and parallel β-strand pair structures than within the anti-parallel β-strand pairs.

Another way to assess the descriptiveness of SCN0s of protein structure is with Relative Contact Order (RCO), which was proposed as a complexity measure of a protein's native topology [7, 8]. RCO = CO/$N$ where CO is the average sequence separation of contacts, and $N$ is the number of residues. When RCO is calculated with native shortcut edges only (SCN0_RCO), we find that it is able to classify proteins by their secondary structure composition (Fig. S7). α-rich proteins tend to have smaller SCN0_RCO values than β-rich proteins. This tendency bodes well for SCN0 as a graph object with which to study protein folding.

**3.2 Native SCN based structural metrics and protein folding kinetics**

Since the serendipitous discovery of the significant and strong negative correlation between RCO and folding rates of small two-state proteins, there has been tension between the relative importance of protein chain length (size) and protein structure to predict protein folding rates [7, 8]. This tension arose, in no small part, due to the failure of RCO and the success of CO to capture non-two-state protein folding kinetics. A resolution to this tension, for the purpose of predicting folding rates from linear extrapolation, is to combine the two determinants into one metric [51, 52]. In contrast, we use this tension to scout for protein folding relevant structural descriptors which are size independent, i.e. metrics that strongly (< -0.3 or > 0.3) correlate with folding kinetic data, but weakly (> -0.3 and < 0.3) correlate with protein size[1]. The goal here is not to maximize linear correlation coefficients, but to identify an effective single-value structural descriptor for folding irrespective of protein size.

Four groups of metrics based on number of edges, clustering coefficient, CO and RCO are explored, with PRN0 edges and with SCN0 edges, on three distinct protein kinetic datasets. lnCO is the average natural logarithm of sequence separation of contacts [53]: $\text{lnCO} = \frac{1}{M} \sum_{i}^{M} \ln(s_i)$ where $M$ is the number of contacts considered, and $s_i$ is the sequence separation of the $i^{\text{th}}$ contact. SCN0_lnCO is lnCO computed on native shortcuts only. lnRCO = lnCO / ln($N$) [53].

*Folding rate of two-state and non-two-state proteins*

The Zou-Ozkan dataset (Table S2) comprises the folding rates ln($k_f$) of 55 two-state proteins of diverse secondary structure (Fig. S8 top-left). The Kamagata dataset [33] comprises folding rates to the native-state log($k_N$) for 23 non-two-state proteins, and unfolding rates log($k_U$) for 19 two-state proteins (Fig. S8 bottom-left). The logarithm of rate constants for the formation of intermediates log($k_I$) for 10 non-two-state folders is also reported in [33]. But given the small size of this dataset, and the very strong correlation between log($k_I$) and log($k_N$) (Fig. S8 bottom-right), we focus only on log($k_N$) for non-two-state proteins.

Of the metrics explored, we find SCN_lnCO to be the only one that best fits our abovementioned criterion for both datasets. SCN_lnCO is significantly correlated with the folding rates of both two-state and non-two-state folders, separately and when combined (the first three plots in Fig. 4, Table 3). It is interesting that the SCN_lnCO is significantly correlated with $N$ for the Zou-Ozkan dataset, but it is not so for the smaller Micheletti dataset, which also considers only two-state folders (the forth plot in Fig. 4, Table 3). Compared with the RCO-related metrics, the CO-related metrics show less sensitivity to change in folding behavior. Aiming for a folding rate determinant that can accommodate mixed folding behaviors is ideal since it not only simplifies matters, but there may not be a hard divide between two-state and non-two-state folding [52].

Another group that appears impervious to change in folding behavior, but is strongly correlated with folding rates, is the edge-related metrics. This group does not satisfy our criterion of a size-independent structural descriptor because its member metrics are strongly correlated with $N$. However, of the four edge-related metrics considered, the number of long-range shortcuts |SCLE|, stands out as the best and most consistent determinant in the two-state, the non-two-state and the combined folding kinetics datasets. The strong negative correlation between |SCLE| and two-state folding rates is noteworthy because it shows how two-state folding kinetics is also dependent on $N$, by way of the strong correlation between |SCLE| and $N$. Unlike the non-two-state folding rates, the two-state folding rates correlate with (raw) $N$ only weakly (the first two plots in Fig. 4).

Another metric that is consistently indifferent to changes in protein size is C_SCN. However, it is not a significant determinant of folding rates for non-two-state proteins, and only a weak, but still significant, determinant of folding rates for two-state proteins (Table 3). C_SCN outperforms C_PRN, which is not a significant determinant of protein kinetics for any dataset, because C_SCN is more sensitive to differences in protein secondary structure than C_PRN (Fig. S9).

---

[1] These ranges of correlation coefficient values are a rough guide only. Consideration is also given to p-values.

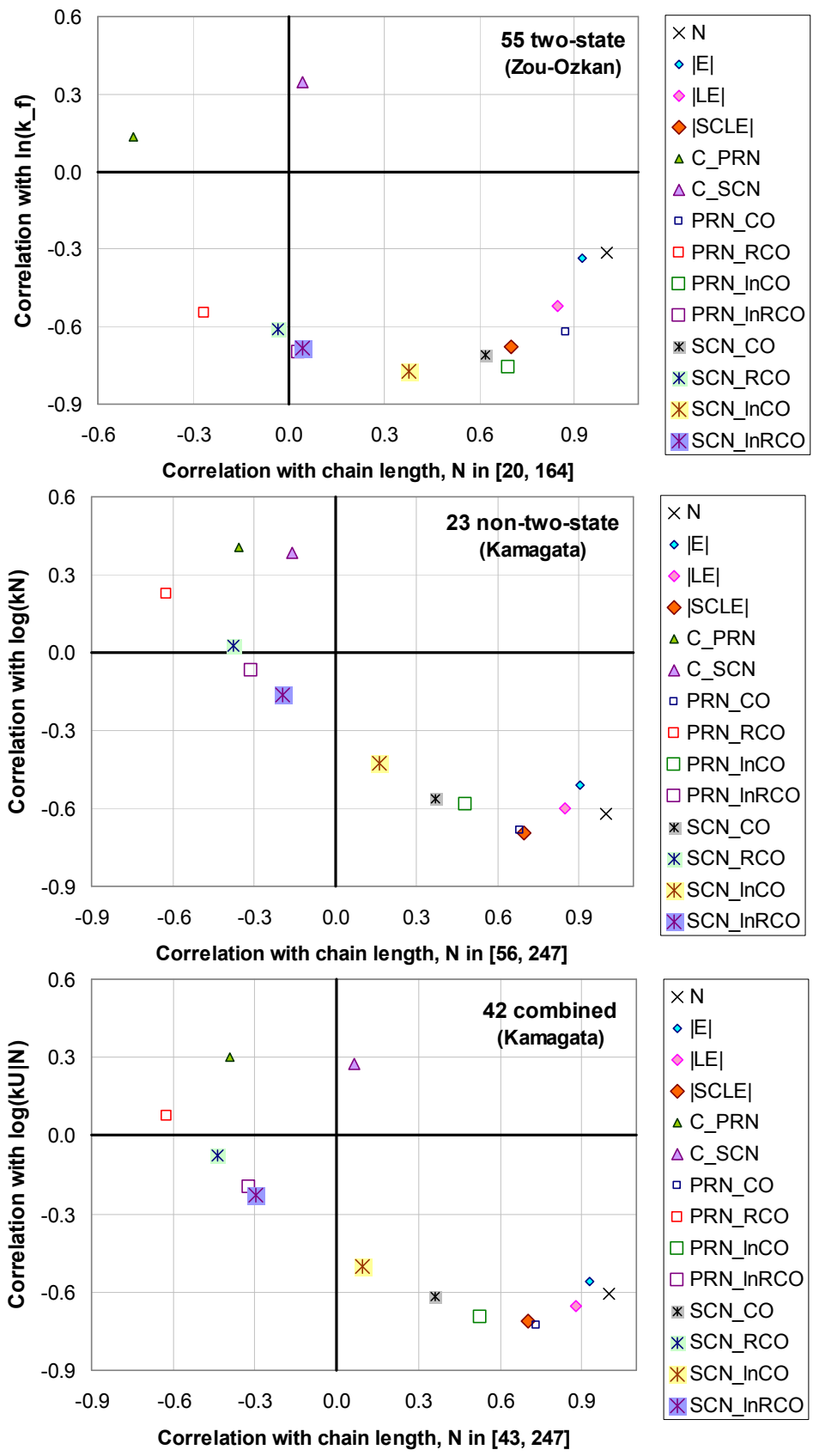

**Fig. 4** Four plots showing how the 14 *native-state* metrics correlate with protein size *N*, and folding kinetics extracted from three datasets.

The edge-related metrics |E|, |LE| and |SCLE| are, respectively, number of PRN edges, number of long-range PRN edges and number of long-range shortcut edges.

The clique-related metrics C_PRN and C_SCN are the clustering coefficient of a PRN and the clustering coefficient of a SCN, respectively.

The Contact Order (CO) and the Relative Contact Order (RCO) related metrics come in two flavors; they depend on whether the set of edges used to compute a metric come from a PRN or a SCN.

The lnCO and lnRCO metrics are defined in the text and they represent the logarithm form of CO and RCO respectively.

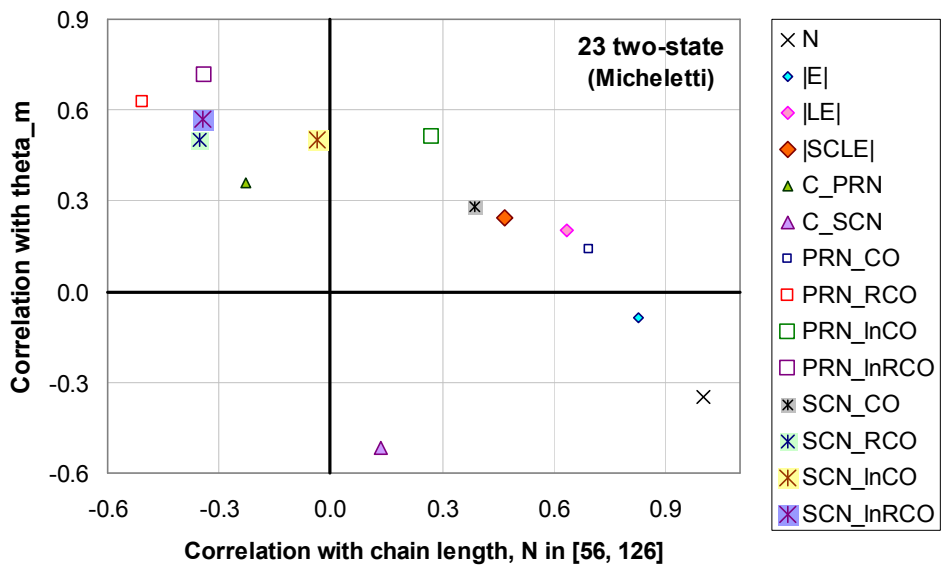

**Table 3** Pearson correlation coefficients (cor) and p-values of three SCN0 based metrics shown in Fig. 4, against protein size *N*, and protein kinetic data.

|  | SCN_lnCO | | C_SCN | | C_PRN | | |SCLE| | |
|---|---|---|---|---|---|---|---|---|
|  | cor | p-value | cor | p-value | cor | p-value | cor | p-value |
| *N* | 0.3785 | 0.0044 | 0.0419 | 0.7615 | -0.4885 | 0.0002 | 0.7008 | 0 |
| Two-state folding rate | -0.7742 | 0 | 0.3447 | 0.0010 | 0.1368 | 0.3191 | -0.6785 | 0 |
| *N* | 0.1607 | 0.4639 | -0.1612 | 0.4624 | -0.3577 | 0.0938 | 0.6965 | 0.0002 |
| Non-two-state folding rate | -0.4248 | 0.0433 | 0.3846 | 0.0699 | 0.4051 | 0.0552 | -0.6940 | 0.0002 |
| *N* | 0.0939 | 0.5543 | 0.0633 | 0.6903 | -0.3933 | 0.0010 | 0.7023 | 0 |
| Combined (un)folding rate | -0.5039 | 0.0007 | 0.2743 | 0.0787 | 0.3033 | 0.0509 | -0.7110 | 0 |
| *N* | -0.0350 | 0.8741 | 0.1333 | 0.5444 | -0.2291 | 0.2930 | 0.4678 | 0.0244 |
| Transition-state placement | 0.5009 | 0.0149 | -0.5158 | 0.0117 | 0.3606 | 0.0910 | 0.2443 | 0.2612 |

The RCO-based metrics can be significantly correlated with *N*. So too other metrics, such as C_PRN, that seem size independent only because *N* is a denominator in their calculations. A lesson here is correlations need to be evaluated carefully with reference to the dimensions of a dataset. A simple division by *N* may not fully neutralize the effects of *N* on a metric, and may even over-correct it.

*Transition-state placement*

The Micheletti dataset [14] comprises transition-state placement data (theta_m) for 23 two-state proteins. There is a weak negative correlation between theta_m and the folding rates $\ln(k_f)$ of these two-state folders (Fig. S8 top-right). Given that a larger theta_m value reflects a transition-state that is more similar to the native-fold in terms of the burial of hydrophobic surface areas, this negative correlation is counter-intuitive, but not unusual. A negative correlation between these two variables was reported in [7] for their dataset.

Both SCN_lnCO and C_SCN are good size-independent determinants for theta_m (the forth plot in Fig. 4, Table 3). These two significant correlations between structural descriptors of proteins and theta_m are noteworthy because they reaffirm the influence of native-state topology on the (native-like) transition-state of two-state folders [54]. The sign of their correlation coefficients are flipped since theta_m and $\ln(k_f)$ are inversely related (Fig. S8 top-right). Cliquishness in a network implies a transitive dependency between contacts: the contact probability between two nodes which are already in contact with a common node is high. If folding rate is viewed as the rate at which native contacts form, then a positive correlation between folding rate and cliquishness follows. But this 'cooperative formation of native interactions' [14] is a statistically significant hindrance to transition-state placement.

We suspect that the inverse relationship between cliquishness of SCN0 and transition-state placement is related to the principle of minimizing entropy loss [45]. Early loop formation promotes early and possibly premature loss in configurational entropy, and so contacts in transitions-state ensembles tend to be those that minimize entropy loss given the structured segments already present [51]. Residue burial is positively correlated with node degree, but only weakly negatively correlated with node clustering (Fig. S10). Residues forming hydrophobic cores and those identified as key folding sites by various methods such as evolutionary conservation, rigidity analysis and hydrogen exchange, have relatively large PRN0 node degree and small PRN0 node clustering (data not shown). Presumably, these key folding residues are those implicated in transition-state placement.

Plaxco et. al. observed that transition-state placement is largely independent of size, but related to topological complexity of the native-state [7]. It is no surprise then that the size-related metrics ($N$, $|E|$, $|LE|$, $|SCLE|$) perform worse than the structure-based metrics ($C$, CO, RCO) with theta_m.

*SCN_lnCO and stability*

The SCN0 based metrics: SCN_lnCO, C_SCN and $|SCLE|$, shine in the various situations discussed above. The performance of SCN_lnCO in particular deserves further discussion, as it emerged as the best overall size-independent structural descriptor for protein folding kinetics. This finding has three important implications. First, it reinforces the relevance of the native-state topology hypothesis to folding kinetics of both two-state and non-two-state proteins, and thus helps to build a basis for a unified theory of globular protein folding. Kamagata et. al. [33] reached a similar conclusion, but their analysis is adulterated with size concerns, and depends on a small set of highly correlated data (Fig. S8 bottom-right). Second, by being significantly correlated to both folding rates and transition-state placement, SCN_lnCO reaffirms the influence of native-state topology on the transition-state of two-state proteins, and upholds the unified view of folding mechanisms for two-state proteins [54]. Third, it reconciles CO with polymer theory which theorizes that the entropy associated with the formation of a (long, length > 5) loop is related to the logarithm of its length [6]. If the physical basis of folding rate dependence on CO is to be explained by the loss of configurational entropy as a result of loop formation, i.e. the loop-closure principle [45, 53], then lnCO is more appropriate than CO as a single-value structural descriptor of proteins for folding kinetics.

CO intends to capture the relative difficulties introduced by local and non-local contacts to protein folding [6, 7]. But by applying a logarithm function to the sequence separation of edges, the differences get "smoothed out" in lnCO [2]. Does this mean that SCN_lnCO reduces topological complexity? One way to look at the numerator of SCN_lnCO is as a "dynamic" topological descriptor, similar in concept to ECO (Effective Contact Order) [45, 53]. As a protein folds, contacts which are distant in a completely stretched out sequence, gradually become less distant with the help of contacts already formed. The denominator of SCN_lnCO is the number of shortcuts $|SC|$, which is strongly and positively correlated with $N$. Thus, SCN_lnCO can still be associated with two of the three major non-sequence specific factors of folding kinetics -- topology and length.

The third major non-sequence specific factor is stability. If folding rates reflect the barrier that is surmounted by a protein sequence to enter the well of its native-state in the folding landscape, stability values reflect the depth of this well. None of the metrics we explored produced a significant correlation with the stability data (experimentally determined free-energy to unfold a protein) provided in [45] for 23 two-state folders[3]. The best (Pearson cor.= 0.5130, p-value= 0.0174) metric, after excluding "outliers" 1URN and 1LMB, is C_SCN ($C_{SCN0}$). This positive correlation is understandable since the stability values provided in [45] show some correlation with secondary structure, and C_SCN is sensitive to secondary structure (Fig. S9). The positive relationship between C_SCN and fold stability also makes intuitive sense: the third contact that completes a triangle (the smallest clique possible) reinforces the two contacts already present, and supplies an alternative path of communication amongst the three nodes should either of the original two contacts fail. In this sense, C_SCN is a more complete structural descriptor of two-state proteins for folding kinetics than SCN_lnCO.

---

[2] A simple example: $128 - 2 = 126$, but $\log_2(128) - \log_2(2) = 6$.
[3] 1AYE is excluded because we could not determine the applicable residue range.

The stability values are also significantly positively correlated (Pearson cor.= 0.4531, p-value= 0.0299) with the number of native short-range shortcuts, |SCSE|, but, somewhat surprisingly, not significantly correlated with the number of native long-range shortcuts, |SCLE|. Perhaps the contribution of SCLE to stability has to be evaluated in the context of their interplay with SCSE. Exclusion of SCLE reduces C_SCN, more pronouncedly for proteins with β-sheets, and SCLE produce negligible or no cliques themselves. Better insight into the influence of stability on protein kinetics and its relationship to native-state topology may be achieved with more refined stability measurements that discriminate between sources or stages of stability: stability due to secondary structure, stability of the nucleus, and stability of tertiary folds due to long-range contacts.

### 3.3 Folding pathways of canonical proteins

The folding pathways identified by RBC with the $C_{SCN0}$ rule and the **E** rule (section 2.8) are summarized in Table 4. A unique positive integer is associated with each pair of parentheses to record the fold order unambiguously. Elements of SSUs in smaller numbered parentheses combine earlier. For example, $(((β1\ β2)_1\ α1)_3\ (β3\ β4)_2)_4$ describes the following sequence of fold events: (i) strands β1 and β2 form SSU(β1 β2), (ii) β3 and β4 pair up to form SSU(β3 β4), (iii) the helix α1 and the first β-hairpin form SSU(β1 β2 α1), and (iv) finally the second hairpin and SSU(β1 β2 α1) form the complete native Conformation SSU(β1 β2 α1 β3 β4).

**Table 4** $C_{SCN0}$ and **E** protein folding pathways for the 12 two-state proteins examined. Yang & Sze [10] folding pathways are provided for reference where available. 'same' denotes that the pathway is identical to the one in the previous column on the left.

| PDB ID | $C_{SCN0}$ folding pathway | **E** folding pathway | Yang & Sze folding pathway |
|---|---|---|---|
| 1BDD | $(α1\ (α2\ α3)_1)_2$ | same | same |
| 2IGD | $((β1\ β2)_2\ (α1\ (β3\ β4)_1)_3)_4$ | same | Not applicable. |
| 1GB1 | $((β1\ β2)_2\ (α1\ (β3\ β4)_1)_3)_4$ | same | $(((α1\ (β3\ β4)_1)_2\ β1)_3\ β2)_4$ |
| 2PTL | $(((β1\ β2)_1\ α1)_3\ (β3\ β4)_2)_4$ | same | $(((β1\ β2)_1\ (β3\ β4)_2)_3\ α1)_4$ |
| 1MHX | $((β1\ β2)_1\ (α1\ (β3\ β4)_2)_3)_4$ | $(((β1\ β2)_2\ α1)_3\ (β3\ β4)_1)_4$ | $((β3\ (β4\ (β1\ β2)_1)_2)_3\ α1)_4$ |
| 1MI0 | $((β1\ β2)_1\ (α1\ (β3\ β4)_2)_3)_4$ | same | $((β3\ (β4\ (β1\ β2)_1)_2)_3\ α1)_4$ |
| 2CI2 | $((β1\ α1)_3\ ((β2\ β3)_1\ β4)_2)_4$ | $(β1\ ((α1\ (β2\ β3)_1)_2\ β4)_3)_4$ | same |
| 1SHG | $(β1\ ((β2\ (β3\ β4)_1)_2\ β5)_3)_4$ | $((β1\ β2)_1\ ((β3\ β4)_2\ β5)_3)_4$ | Not applicable. |
| 1SRM | $(β1\ (((β2\ β3)_1\ β4)_2\ β5)_3)_4$ | $((β1\ β2)_2\ (β3\ (β4\ β5)_1)_3)_4$ | Not applicable. |
| 2KJV | $(((β1\ α1)_2\ (β2\ β3)_1)_3\ (α2\ β4)_4)_5$ | $(((β1\ α1)_2\ β2)_4\ (β3\ (α2\ β4)_1)_3)_5$ | Not applicable. |
| 2KJW | $((((β3\ α2)_2\ (β4\ β1)_1)_3\ α1)_4\ β2)_5$ | $((β3\ (α2\ β4)_1)_3\ ((β1\ α1)_2\ β2)_4)_5$ | Not applicable. |
| 2F4K | $((α1\ α2)_1\ α3)_2$ | Not applicable. See section 3.5. | Not applicable. |

The plots in Fig. S11 show the energy of Conformations on $C_{SCN0}$ and **E** folding pathways against the fraction of shortcut edges. The $C_{SCN0}$ and **E** folding pathways are identical for 1BDD, 2IGD, 1GB1, 2PTL and 1MI0, but diverge for 1MHX, 2CI2, 1SRM, 1SHG, 2KJV and 2KJW. With the exception of 1MHX and 1MI0 for which energy increases in the final step(s), the $C_{SCN0}$ and the **E** folding pathways are energetically favourable: energy decreases with each fold step as the search approaches the native conformation. This behavior is consistent with the funnel energy landscape model for small proteins.

1BDD is the B domain of *Staphylococcal* protein A. Its secondary structure comprises three helices (Fig. 1). 1BDD is an almost symmetric protein, which prompts questions about how the symmetry is broken during (un)folding, and the possibility of multiple highly populated (un)folding pathways under different conditions [55]. While experiments and simulations concur that the middle α-helix (α2) is the most stable structure, there is disagreement about the sequence of fold events involving the first and third α-helices in the transition state [55, 56, 57]. The $C_{SCN0}$ folding pathway for 1BDD is identical to the folding pathway predicted by Yang & Sze, and by COMMA [26]. This pathway is the dominant unfolding pathway passing through TS2 under condition of high temperature or high denaturant concentration [55]. A non-random choice between the two possible folding routes for 1BDD could not be made with a simple count of native shortcuts since both Conformations (1H 3H)(5H) and (1H)(3H 5H) have the same number of shortcuts (Table 1). The alternative folding pathway that passes through TS1 in [55], which is supported by experiments but hard to simulate, could be selected using a different rule since Conformation (1H 3H)(5H) boasts more native PRN edges and also more hydrogen bonds than Conformation (1H)(3H 5H). However, these alternative rules do not generalize well; they break down for the next four proteins.

2IGD and 1GB1 are two PDB entries for the B1 immunoglobulin binding domain of *Streptococcal* protein G. They are composed of one α-helix packed against a four-stranded β-sheet comprising two β-hairpins (Fig. S12). 2IGD and 1GB1 have identical $C_{SCN0}$ and **E** folding pathways, and their respective $C_{SCN0}$ and **E** folding pathways coincide. All the folding routes in Table 4 for 2IGD and 1GB1 begin with the formation of the C-terminus β-hairpin (β3-β4); this is consistent with literature [9]. The $C_{SCN0}$ rule pairs β3 with β4 as the first fold event in 49 of the 60 1GB1 NMR models.

2PTL is the B1 immunoglobulin binding domain of *Peptostreptococcal* protein L. Like protein G, it is also composed of one α-helix packed against a four-stranded β-sheet comprising two β-hairpins. Its $C_{SCN0}$ and **E** folding pathways coincide, and agree that folding begins with the formation of the N-terminus β-hairpin (β1-β2), in-line with literature [9, 58]. The $C_{SCN0}$ rule identifies (β1-β2) as the first fold event in six of the 20 2PTL NMR models.

1MHX and 1MIO are crystallized structures of the NuG1 and NuG2 variants of protein G [59]. These variants were made by searching for a replacement sequence in the first (N-terminus) β-hairpin of protein G that would increase its stability, and cause a switch to protein G's dominant folding pathway [60]. All except one of the folding routes in Table 4 for 1MHX and 1MI0 agree that folding begins with the formation of the N-terminus β-hairpin (β1-β2). The exception is 1MHX's **E** folding pathway which proposes the second β-hairpin.

Results from the set of four structurally similar proteins just discussed support the applicability of $C_{SCN0}$, a structural metric based on native-state topology, to the protein folding pathway problem. In all four cases, $C_{SCN0}$ correctly identified the expected initial reaction point. The correct initial reaction point for these four proteins could not be reproduced simultaneously with alternative rules such as the number of SCN0 edges, the number of PRN0 edges, $C_{PRN0}$, the number of hydrogen bonds, a random subset of PRN0 edges, SSE communication strength [26], β-hairpin contact density or β-hairpin energy (data not shown).

The fold events on the $C_{SCN0}$ pathways for the above four proteins are quite regimented. First, the two β-hairpins are formed one after the other, then the α-helix joins one of the β-hairpins and finally the other β-hairpin is included. The α-helix docks onto the β-hairpin that forms first in the cases of 2IGD, 1GB1 and 2PTL. But for both 1MHX and 1MIO, the α-helix still docks onto the second β-hairpin, as in the wild-type protein G. The two-strand-helix structural combination has been recognized as a preferred folding nucleus starter in α/β proteins [61], and is probably the smallest possible nucleation motif for globular proteins [37]. The β1-β2, α1 and β3-β4 substructures are requisite folding elements on low ECO routes for proteins G and L [45], and coincide also with substructures in the formation order suggested by rigidity analysis [62].

In contrast, the fold events on Yang and Sze's pathways for this set of four proteins are more diverse, and all of them involve an intermediate Conformation that is infeasible in our RBC model. Support for early pairing of the first and fourth β-strands to form a three-stranded β-sheet in the transition state have been documented for both proteins G and L [9, 20, 22]. There is a proclivity for early coupling of N- and C-termini in $C_{SCN0}$ pathways if the formation of N-C SSUs is permitted (Table S5).

In general, we observe that RBC with $C_{SCN0}$ prefers to combine adjacent β-strands early since β-strands have zero $C_{SCN0}$ individually, and therefore their combination (formation of a β-hairpin) can only affect the fitness of a Conformation in a non-decreasing manner. In contrast, long α-helices tend to have large $C_{SCN0}$ values, and their inclusion into a SSU maybe delayed since this combination could potentially decrease the fitness of a Conformation.

2CI2 is the Chymotrypsin Inhibitor 2 protein. It has one α-helix, a reactive site loop and a six-stranded β-sheet (which is simplified to four in some models as we do in RBC) (Fig. S13). Its **E** folding pathway coincides with Yang and Sze's folding pathway, and the low fold order of SSU(α1 (β2 β3)), reflects the early formation of the major hydrophobic core by its nucleus residues [63]. The $C_{SCN0}$ folding pathway suggests the early formation of the three stranded β-sheet SSU((β2 β3) β4) instead. Nonetheless, all the 2CI2 pathways in Table 4 agree with results from experiments and simulations that folding is instigated by the β2-β3 substructure (or β3-β4 if dealing with the six-stranded β-sheet model) [21, 35]. In the six-stranded β-sheet model, the parallel β3-β4 structure is partially formed in the transition state [35], while strands β1, β5 and β6 (or β1 and β4 in the four-stranded β-sheet model) appear completely unstructured in the transition state [63].

1SHG and 1SRM are the src-homology 3 (SH3) domains of the α-spectrin and Src tyrosine kinase proteins respectively. Both comprise a five-stranded β-sheet (Fig. S14). The folding pathways of these two structurally homologous proteins are expected to be the same since their folding transition states closely resemble each other [34]. For both SH3 domains, the distal β-hairpin (β3-β4) is the best formed structural element in the transition state [34, 64]. And indeed, except for the first fold event, the $C_{SCN0}$ folding pathways of 1SHG and 1SRM are identical and SSU(β3, β4) has a low fold order. $C_{SCN0}$ identified β2-β3 and β3-β4 as the respective initial folding reaction point in five and 13 of the 20 1SRM NMR models. The same cannot be said for their **E** folding pathways which differ from each other in the crucial first and second fold events. The **E** folding pathways are less plausible as they involve either the N- or the C-terminus β-strands early on, both of which are largely unstructured in the transition state ensemble [34, 64].

The $C_{SCN0}$ folding pathways of 1SHG and 1SRM identify the early formation of the β-sheet composed of β2, β3 and β4. This is in accord with experiments and computer simulations [34, 64]. The formation of the central three-stranded β-sheet is supported by the diverging turn (just before β2), the n-src loop (between β2 and β3), and the distal-loop (between β3 and β4). These turn and loop regions are highly populated in computer simulations of the src SH3 domain [64]. The protein engineering method (φ-analysis) reveals that the diverging turn and β2 of the src SH3 domain are partially ordered in the transition state and interact with the distal β-hairpin (β3-β4) to form the central β-sheet [64]. A hydrogen bond network [65] and a network of interactions between hydrophobic residues [66] have been detected between the diverging turn and the distal β-hairpin in the folding transition state. Mutations on the central β-sheet significantly decreased the folding rate of the src SH3 domain [64]. Of the three loop/turn elements supporting β2-β3-β4 in SH3 domains, the distal loop is the stiffest [67], and its disruption retards α-spectrin SH3 folding rate the most [68].

2CI2 is unique in the set of proteins examined in that its parallel β-strand pair of some length is the initial folding reaction point. Formation of β-hairpins (anti-parallel β-strand pairs) in folding transition states is attractive since native contacts can be satisfied with minimal loss in chain entropy [58]. Native contacts with smaller sequence distances can be satisfied before native contacts with longer sequence distances in a zipper-closure like motion. However, long-range native contacts are needed to stabilize parallel β-strand pairs, and there is no apparent mechanism, such as a zipper, to build up these long-range native contacts gradually. But Nature has a varied toolkit. In Fig. S5 (bottom), we see that 2CI2's parallel β2-β3 strands are supported by many more hydrogen bonds than the β-hairpin of 2IGD (Fig. S5 middle).

There are no shortcut edges in the reactive site loop structure of 2CI2, but shortcut edges are present in the turn structure of 2IGD, and in the RT-loop of 1SRM (Fig. S4). The shortcut edges in the RT-loop exhibit a less regular pattern than those within the primary secondary structure forms. CI2's reactive loop structure is a "permissive" location; it is able to tolerate cleavage and accommodate insertion of residues without significantly affecting CI2's stability, folding pathway or folding rate constant [47, 69]. In contrast, disruption to any of the three central loops or turns of the α-spectrin SH3 domain, i.e. RT loop, n-Src loop or distal-loop, affects both stability and kinetic behavior of the circular permutant relative to wild-type, but not the final 3D structure [68].

2KJV is the 30S ribosomal protein S6 of *Thermus thermophilus*. 2KJW is its p54-55 circular permutant. 2KJV and 2KJW have the same tertiary structure, even though the SSEs of 2KJV are rearranged in 2KJW [4], but they have different folding pathways [37]. S6 has two 'structural foldons' which affords it two alternative points to initiate folding [37]. S6's two structural foldons also work cooperatively on the same folding pathway due to their overlap, which triggers the formation of one when the other becomes structured. On 2KJV's $C_{SCN0}$ folding pathway, elements of the first structural foldon (β1, α1 and β3) are involved in the first two fold events, and the first structural foldon is formed by the third fold step. On 2KJW's $C_{SCN0}$ folding pathway, elements of the second structural foldon (β1, α2 and β4) are involved in the first two fold events, and the second structural foldon is formed by the third fold step. Folding begins with the pairing between β2 and β3 for 2KJV, and the pairing between β1 and β4 for 2KJW, on the respective $C_{SCN0}$ folding pathways. The $C_{SCN0}$ rule chose these same initial reaction points for the other 19 2KJV and 19 2KJW NMR models. In short, the $C_{SCN0}$ folding pathways for both 2KJV and its circular permutant 2KJW are reasonable according to existing experimental and simulation studies [61, 70].

In contrast, the β2-β3 and β1-β4 pairings are respectively the last to occur on the 2KJV and 2KJW **E** folding pathways, whose sequence of fold events split the proteins into the following two SSUs: (β1, α1, β2) and (β3, α2, β4). β2 contributes little to S6's folding [70], and the arrangement of SSEs in 2KJW permits it to be the last SSE involved on the $C_{SCN0}$ folding pathway for 2KJW.

### 3.4 Non-canonical proteins

RBC permits collision between two SSUs only if they are consecutively located on a protein chain (section 2.5). This excludes pairing of N- and C- termini SSEs (without the interleaving SSEs) to compose a feasible SSU. However, such an infeasible SSU could be necessary to accommodate the folding reaction of the four helix bundle ACBP, whose rate-limiting native-like structure (RLNLS) [5] has the N- and C-termini helix segments lie almost parallel to each other, tethered by contacts between eight conserved hydrophobic residues located on the two embracing helixes [71]. The rest of the ACBP polypeptide chain in the RLNLS is sparsely ordered with isolated structured fragments of low stability. This flexibility is presumed to facilitate the development of the RLNLS [72].

An account of the early folding events that could lead up to the RLNLS has been the quest of many subsequent experimental and simulation studies on ensembles of denatured ACBP structures. Three possible ACBP folding pathways have emerged from these studies. Each of these pathway possibilities involves the early formation of a complex comprising one of the following helix combinations: (A) α1, α3 and α4 [71, 72]; (B) α1, α2 and α4 [73] and (C) α2, α3 and α4 [74]. The possible existence of these complexes is supported by the detection of significant native-like and non-native long- and medium-range interactions evidencing long-range (inter-helical) residual structure in the ensemble of denatured structures [74-77].

---

[4] We use 2KJV'2 SSE labels for 2KJW in Table 4. For example, α1 refers to the same helix in both 2KJV and 2KJW, but in RBC, it is 3H for 2KJV and 9H for 2KJW. See Table S3.
[5] The slowest to form but obligate native-like structure for productive folding.

To explore the $C_{SCN0}$ folding pathway propensities for ACBP within RBC, we ran an unrestricted version of RBC (*free*RBC) with the 29 PDB NMR structures for bovine ACBP (2ABD), and the 20 PDB NMR structures for yeast ACBP (1ST7). In *free*RBC, SSEs and SSUs are free to collide with each other, but still only two at a time. The results are summarized in Fig. 5.

The most frequently occurring $C_{SCN0}$ folding pathway for 2ABD with *free*RBC involves complex A (13/29). This is followed by pathways involving complex C (8/29), and complex B (6/29). For 1ST7, the most frequently occurring $C_{SCN0}$ folding pathway produced with *free*RBC involves complex C (10/20). This is followed by pathways involving complex A (6/20), and complex B (4/20). With RBC, there are only two possibilities, and the majority of the $C_{SCN0}$ folding pathways for both 2ABD (20/29) and 1ST7 (17/20) involve complex C. At this high level at least, both *free*RCB and RBC produce reasonable $C_{SCN0}$ folding pathways for ACBP. That is, they clearly eschew pathways involving complex D (α1-α2-α3), which has no literature support. In addition, complex D is missing α4, which is the most stable helix in the denatured state [73], and likely plays a key role in instigating the RLNLS [77].

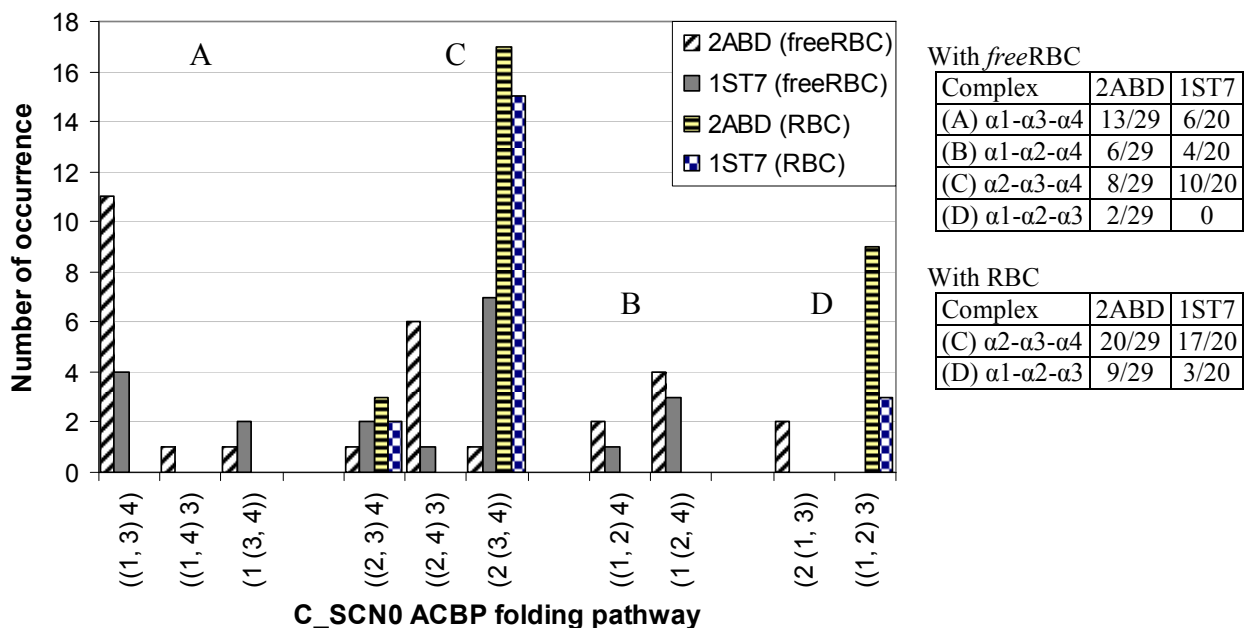

**Fig. 5** $C_{SCN0}$ folding pathway propensities for ACBP with *free*RBC and with RBC, grouped by complex (triplets of segments that take on helical structure in the native state). All possible combinations are explored, and only those that register a non-zero occurrence are reported.

α4 is a member of all three proposed complexes. Residues from the transient α4 helical structure in the denatured state interact with hydrophobic residues in the α2 segment to form important stabilizing interactions for the RLNLS [73, 77]. This observation led to the proposed complex B pathway [73], which also finds support from a strong showing of significant non-random α2-α4 interactions in the ensemble of denatured structures [74, 77]. In Fig. 5, we find at least twice as many complex B pathways that begin with α2-α4 than with α1-α2, and none that initiate with α1-α4. However, it turns out that there is little evidence for the presence of complex B in the unfolded state [74]. Instead, ref. [74] proposes that complex C precedes complex B such that the native α1-α4 interactions in the RLNLS is actually the result of α1 interacting with the α2-α4 part of complex C (α2-α3-α4), thereby forming complex B. For both 2ABD and 1ST7, complex C is more popular than complex B.

Complex A coincides with two of ACBP's three hydrophobic mini-cores [71]. Only one of the $C_{SCN0}$ pathways involving complex A is initiated by a pairing of the N- and C-terminus helices (α1-α4). While the N- and C- terminus regions do make significant non-random contacts in the denatured state, the specific contacts in the folding transition state (RLNLS) mentioned in [71] occur with very low probability in the denatured state [77]. This could be an effect of their large sequence distances [77], and is consistent with the notion that protein folding prefers low ECO routes [45] (which is the argument for the restricted collisions in RBC).

Most of complex A pathways are initiated with α1-α3 (Fig. 5). The difficulty with this initiation point is the absence of α1-α3 contacts in the transition state, and the dearth of significant non-random α1-α3 contacts in the denatured state [77]. The contact map analysis conducted in [77] searched for residue interactions in the ensemble of unfolded structures under several denaturing conditions that have a probability to occur higher than expected in a random coil reference state. Possibly due to its low helical propensity under denaturing conditions [73, 75] and low stability even in the native state, α1 residues make very few such significant non-random contacts with any of the other residues in the denatured state [77]. A similar finding was made in [74]. This lack of interaction between α1 and the other helical segments is problematic for initiating pairs and complexes involving α1 if significant structural features in denatured structures are to hold important clues to early protein folding events, and early folding events ought to influence the subsequent folding pathway.

On the other hand, not all significant interactions in the denatured structure are fold-inducing. There may be persistent non-native interactions in the denatured state that interfere with productive[6] folding, or even cause mis-folding. An example of this is found with α2-α3 non-native interactions [71, 77]. Note that α2-α3 is not a popular initiation site for $C_{SCN0}$ folding pathways. The mostly frequently occurring initiation site for complex C, which involves both α2 and α3, is α3-α4. In the native state, α3 and α4 are not in contact with each other; however many persistent non-native α3-α4 interactions are detected in the denatured state [74, 77]. In the native structure, α2 is flanked by α3 and α4 on each side. Complex C is held together by native-like long-range links in the unfolded state [74].

To summarize, if the sequence of folding events proposed in [74] holds, i.e. that ACBP folding proceeds with early formation of complex C, and given the inter-helical contact probabilities observed in [74] and [77] which are qualitatively at least in agreement with each other, then our RBC model, even with its collision restrictions, is applicable to the expression of ACBP's folding reaction.

Top7 (PDB id: 1QYS) is a 92-residue α/β de novo protein composed of two α-helices and a five-strand anti-parallel β-sheet. 1QYS exhibits non-cooperative and non-two-state folding kinetics [38]. Its folding passes through a very stable intermediate stage where the C-terminus fragment (CFr) comprising one α-helix (α2) and three β-strands (β3, β4, β5) is well formed, and the N-terminus fragment remains disordered most of the time [38].

The $C_{SCN0}$ folding pathway for 1QYS is ( (β1 β2)$_1$ (α1 (β3 (α2 (β4 β5)$_2$ )$_3$ )$_4$ )$_5$ )$_6$. Except for the first step, RBC folding action for 1QYS progresses sequentially from the C-terminus. The CFr is formed by the forth fold step. The CFr SSU (β3 α2 β4 β5) does not form if early N-C termini coupling is allowed (Table S5), nor if fitness of SSUs in a Conformation is evaluated in isolation from each other (Eq. 2 in Note S2).

Table S4 gives four examples where it is preferable to include shortcuts between multi-SSE SSUs when calculating Conformation fitness (as we have done so far with Eq. 1). However, such interactions could prove "too cooperative" for large multi-domain proteins [40]. An *ad-hoc* remedy is to impose a threshold to limit these interactions. The threshold is set to 100 residues, the characteristic maximum length of small two-state proteins. This means that fitness of a newly minted (multi-SSE) SSU $g$ is evaluated with all other existing multi-SSE SSUs $h_1, h_2, ... h_q$ in the first term of Eq. 1 provided that the leftmost and rightmost ends of the protein sequence covered by $g$ and $h_i$ is less than 100 residues in length. The multi-SSE SSUs that lie outside this boundary are evaluated in isolation, just like single-SSE SSUs in the second term of Eq. 1. We use this modified Eq. 1 to identify the $C_{SCN0}$ folding pathway for 7DFR.

---

[6] To have a right structure in a timely manner so as to be functional.

DHFR (*dihydrofolate reductase*) is a much studied model protein with multiple PDB entries. We use the crystallized DHFR structure from *E. coli* 7DFR, which according to the SSE delineation[7] in [39] has four α-helices and eight β-strands. DHFR has two structural domains: ABD (adenine binding domain) between residues 38 and 88, and a discontinuous domain (DD) spanning residues 1...37 and 89...159 [78]. The two domains need to cooperate with each other during folding as they are independently unstable [39, 40].

Evaluation of the $C_{SCN0}$ folding pathway for 7DFR (Fig. 6) relies on the analysis in [39], which compared the stabilities of individual protein fragments identified via experimental cuts (E-cut) [78] and via computational cuts (C-cut) [79], and concluded that computational cuts tend to carve out more stable fragments. This conclusion is good news for us since SSUs on the $C_{SCN0}$ folding pathway for 7DFR align more readily with the C-cut fragments. SSU(16S 18S 20S) forms by step 3, and aligns perfectly with the C-cut fragment 104...159, which is more stable than the E-cut fragment 89...159. SSU(7S 9S 11H 13S 14H), which forms by step 6, fits into the C-cut fragment 35...103. SSU(7S 9S 11H) is part of ABD, and forms by step 4.

$$((\beta 5\ \alpha 1)_{10}\ ((\beta 3\ \alpha 2)_8\ (((((\beta 2\ \beta 1)_2\ \alpha 3)_4\ \beta 4)_5\ \alpha 4)_6\ (\beta 6\ (\beta 8\ \beta 7)_1)_3)_7)_9)_{11}$$

$$((0S\ 2H)_{10}\ ((4S\ 5H)_8\ (((((7S\ 9S)_2\ 11H)_4\ 13S)_5\ 14H)_6\ (16S\ (18S\ 20S)_1)_3)_7)_9)_{11}$$

**Fig. 6** Top: The $C_{SCN0}$ folding pathway for 7FDR, given in two forms. Bottom: An illustration with color-blocking of the stages in the $C_{SCN0}$ folding pathway discussed in the text, and the residue ranges of fragments produced by experimental cuts and by computational cuts.

Under simulation, both the E-cut fragment 37...88 (ABD) and the C-cut fragment 35...103 undergo significant unfolding over time [39]. There are no SSUs on the $C_{SCN0}$ folding pathway for 7DFR that align perfectly with either of these unstable fragments. Instead, RBC combines SSUs from ABD and the C-terminus of DD in three steps starting with SSU(7S 9S 11H 13S 14H 16S 18S 20S) in step 7, and ending with SSU(4S 5H 7S 9S 11H 13S 14H 16S 18S 20S) in step 9. This step 9 SSU is a stable fragment; without residues 1...36, the two domains can associate and form a stable conformation [39].

The E-cut fragment 1...36 and the C-cut fragment 5...35 both experience fast unfolding. SSU(0S 2H) is the penultimate multi-SSE SSU, which means it exerts little influence over the $C_{SCN0}$ folding pathway for 7DFR.

---

[7] The exact residue ranges differ slightly from DSSP or current PDB delineations. Our results for 7DFR are sensitive to SSE delineations.

To summarize, the proposed $C_{SCN0}$ folding pathway for 7DFR has attributes to recommend its reasonableness in terms of fragment stability. Without the 100-residue limit modification to Eq. 1, SSU(0S 2H) forms earlier and produces SSU(0S 2H 4S 5H) in the penultimate step, which splits ABD. Therefore, the 100-residue limit does have a significant positive effect for 7DFR. Will this method generalize to other larger multi-domain proteins? It is premature to think so at this time; certainly more investigation is needed to answer either way definitively. Nonetheless, this approach has several advantages: (i) the so-called folding domains need not be known in advance, (ii) the self-folding steps of a protein sequence is identified from the bottom-up instead of from the top-down [80], and (iii) the guiding role of structure in the folding process is explicit.

### 3.5 SCN analysis of MD generated non-native structures

Discussions and references in the previous sections intimate the possibility of detecting early productive folding events by examining ensembles of denatured structures. This approach has not only been illuminating for ACBP, but also for many other proteins including nearly symmetric folds such as 1BDD [55], and proteins G and L [9, 58, 81]. The effectiveness of this approach, the correlation between native structure contact order and folding rates [54], and the efficacy of native topology to predict folding pathways imply that the influence of a protein's native-state topology extends to its transition and even denatured structures, at least for small two-state proteins [9, 20]. Hence, for $C_{SCN0}$ to be a convincing descriptor of native-state topology to guide folding, its signal needs to be detected in non-native structures as well.

We explore this issue here with the much studied Villin Headpiece subdomain protein (PDB id: 2F4K) and its MD generated structures [41]. 2F4K has three α-helices. Its $C_{SCN0}$ folding pathway (Table 4) begins with SSU(1H 3H). The first and second α-helices are favoured to associate first since situated on them are three conserved phenylalanine hydrophobic core residues (F47, F51 and F58) that are crucial for the stability of the native wild-type villin structure [82]. A SCN0 is constructed for each snapshot configuration taken from 30 MD runs (Note S3). The SCN0 of a snapshot configuration includes only those snapshot shortcuts that are also native shortcuts.

Prima facie, three observations support the use of SCN0 to investigate the MD generated structures. (i) The initial structures (first snapshots in MD runs) have smaller, less cliquish SCN0s than the native structure. This implies that the initial structures are less compact structures than 2F4K. (ii) There is a significant strong positive correlation between the number of native shortcuts and the number of native alpha residues (Fig. S15). Further, the successful runs (those starting with initial structures 4, 7 or 8) have more of each. (iii) The unsuccessful runs explore a significantly (one-sided t.test p-value 3.35E-06) larger set of shortcuts than the successful runs (Table S6). The restriction in conformational sampling indicates that the successful runs have more relaxed configurations.

*Detection of $C_{SCN0}$ bias in non-native structures*

Next, we compute the mean $C_{SCN0}$ for Conformation(1H 3H)(5H) and for Conformation(1H)(3H 5H) of each run. The difference between these two mean $C_{SCN0}$ values per run is significantly (one-sided t.test p-value 0.0057) larger in the successful than in the unsuccessful runs. The average difference between the two mean $C_{SCN0}$ values is 0.2129 (std. dev. 0.0657) in the successful runs, and 0.1302 (std. dev. 0.0970) in the unsuccessful runs.

When the fraction of snapshots in a run where $C_{SCN0}$ (1H 3H)(5H) > $C_{SCN0}$ (1H)(3H 5H) is examined, the fraction is significantly (one-sided t.test p-value 0.0094) larger on average in the successful than in the unsuccessful runs. Overall, the successful runs exhibit a more rapid and steady increase in the cumulative number of snapshots where $C_{SCN0}$ (1H 3H)(5H) > $C_{SCN0}$ (1H)(3H 5H) over simulation time (Fig. 7).

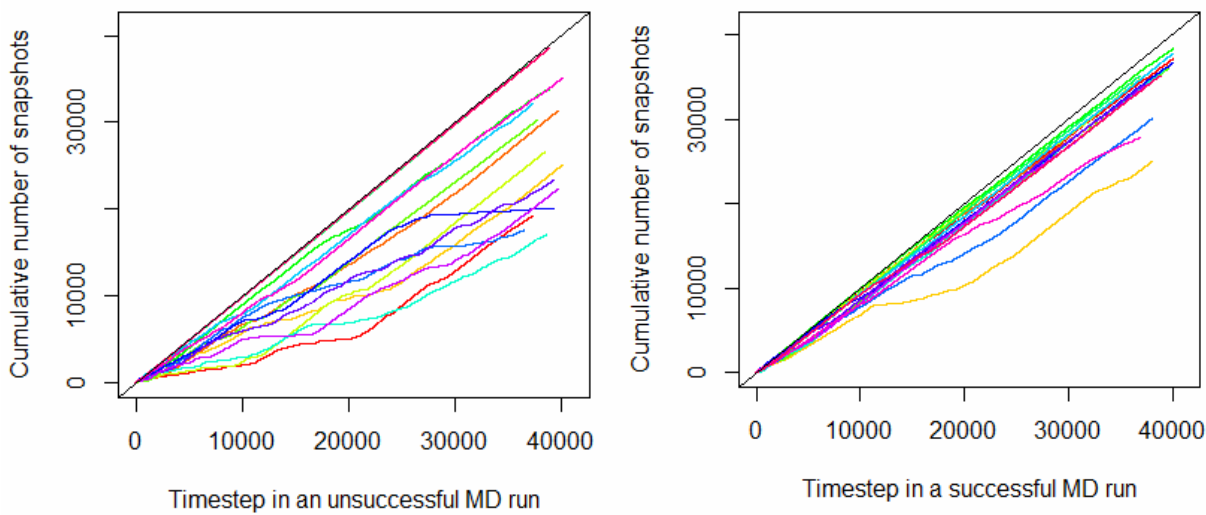

**Fig. 7** Cumulative number of snapshots by MD run where Conformation(1H 3H)(5H) has a larger $C_{SCN0}$ value than Conformation(1H)(3H 5H). Overall, the increase is slower in the unsuccessful runs (left) then in the successful runs (right).

We conclude from these observations that the influence of $C_{SCN0}$ on 2F4K's folding pathway is detectable even in non-native structures. The non-native structures show the same $C_{SCN0}$ bias as the native 2F4K structure. This bias is significantly stronger in the successful runs, and supports the $C_{SCN0}$ folding pathway for 2F4K.

*Fold success factors*

The goal of protein MD simulation is insight into protein folding and function via its movements. What are the critical contacts for folding? How might productive folding be foiled? Here we report three fold success factors for 2F4K suggested by network analysis of shortcuts in its MD generated structures. For this investigation, we use SCN, the shortcut network of a MD snapshot (includes both native and non-native shortcuts) because 2F4K has no native long-range shortcuts. The unique SCNs from the 30 MD runs are classified into five sets: (**A**) appears in **unsuccessful** runs only, (**B**) appears in **successful** runs only, (**C**) appears in **fast-successful** runs only (in at least one r4 or one r7 trajectory and no non-r4 or non-r7 trajectories), (**D**) appears in **slow-successful** runs only (in at least one r8 trajectory and no non-r8 trajectories), and (**E**) appears in **all successful** runs only (in at least one r4 trajectory and one r7 trajectory and one r8 trajectory and no unsuccessful trajectories). Sets C, D and E are a further classification of SCNs in B (Fig. S16). Our global approach to SCN analysis views the MD runs as samplers of the configuration landscape. The previous conclusion holds with $C_{SCN}$, and within this global examination of SCNs (Tables S7A & S7B). Most strikingly, 100% of the E set SCNs have $C_{SCN}$ (1H 3H)(5H) > $C_{SCN}$ (1H)(3H 5H).

*(i) Long-range shortcut edges*

In section 3.2, we saw how (native) |SCLE| is a significant size-dependent determinant of protein folding rate. With the 2F4K MD runs, we have an opportunity to test this influence, albeit with non-native SCLE, in a controlled situation, i.e. on the same protein sequence as it folds. And indeed, the successful runs have significantly fewer SCLE than the unsuccessful runs, and the fast-successful runs have significantly fewer SCLE than the slow-successful runs (Tables S7A & S7B). The E set of SCNs has the fewest SCLE.

Long-range contacts are not impediments to folding per se, and a good dose of them, suitably located, may even be essential to prepare the way for the RLNLS, as is the case with ACBP [74]. Even the wild-type Villin headpiece has a non-local contact (F47-F58) in its hydrophobic core. A characteristic of SCLE in the successful 2F4K MD runs is they occur more frequently between the first and second α-helices (Tables S7A). The proportion of SCLE that connect 1H with 3H is 11% in the unsuccessful runs (A set of SCNs), and 18% in the successful runs (B set of SCNs). This proportion increases to almost 46% in the E set of SCNs.

*(ii) Role of 1H and development of SSU(1H 3H)*

Ensign *et al.* found it difficult to discern how the first α-helix influences 2F4K folding because traces of its structure was found in the initial configurations of both successful (4 & 7) and unsuccessful (0 & 1) runs [41]. This confusion is cleared up somewhat with $C_{SCN0}$ since all initial structures, except for runs 4 and 7, have zero $C_{SCN0}$(1H) (Table S8). In our analysis, the initial structure of a run is captured by its first MD snapshot.

However, since this still does not explain the success of runs 8, we monitored the development of the three α-helices by counting the number of native triangles their SCN0s form over simulation time relative to their total number of triangles in the native structure. $rT_{SCN0}$(H) is the number of triangles formed by SCN0 in H's current structure, relative to the number of triangles in H's native structure. An α-helix is more developed the closer its $rT_{SCN0}$ is to 1.0. The SCN of 2F4K makes four, two and six triangles for 1H, 3H and 5H respectively.

For both unsuccessful and successful runs (all SCN categories A to E), 1H is significantly less developed than both 3H and 5H (Table S9B). The most developed α-helix is 5H for the unsuccessful runs (A), and 3H for the successful runs (B...D). In category E, both 3H and 5H are almost fully developed, but 1H is only 20% developed on average (Table S9A). Taken together, these observations suggest an interesting point. That while a non-zero $C_{SCN0}$(1H) in the initial starting configuration could play a role in the fast folding of 2F4K, it is the development of the second α-helix (3H) during the folding process that appears to assure fold success.

There is a significant positive correlation between $rT_{SCN0}$(1H) and $rT_{SCN0}$(1H 3H) (Table S9C). Of the possible α-helix combinations, $rT_{SCN0}$(1H) is most strongly correlated with $rT_{SCN0}$(1H 3H). This suggests a possible positive influence of the development of 1H on the formation of SSU(1H 3H), which is the RBC suggested initial folding reaction point for 2F4K. This correlation supports the suggestion made by Ensign *et al.* [41] that 1H be mutated to increase its stability to ensure fast folding of 2F4K.

*(iii) Fold inducing shortcuts*

In the ensemble of acid-unfolded ACBP structures, Thomsen et al [75] noticed a subpopulation with a particular set of native-like interactions which have a high propensity to fold. This observation inspired further scrutiny of the E set edges. What are the shortcut edges common to all the successful runs?

The E SCNs cover 77 unique edges, which are depicted in Fig. 8 in contact map form. The three α-helical structures are evident from this contact map. There are only two long-range contacts, (43, 55) and (44, 56), both of which occur between the start of 1H and the start of 3H. 35 of the E edges appear in every E SCN. 96% of the E edges are native. All but one native shortcut is in E; this reinforces the importance of having the right shortcuts for successful folding.

The E edges are more likely to appear in a successful only (B) SCN than in an unsuccessful only (A) SCN, while non-E edges (edges appearing in the other four SCN categories and not in E) are less likely to appear in a B SCN than in an A SCN (Fig. S17). The presence of several E edges in a non-native structure could assist in identifying promising transient structures outside E.

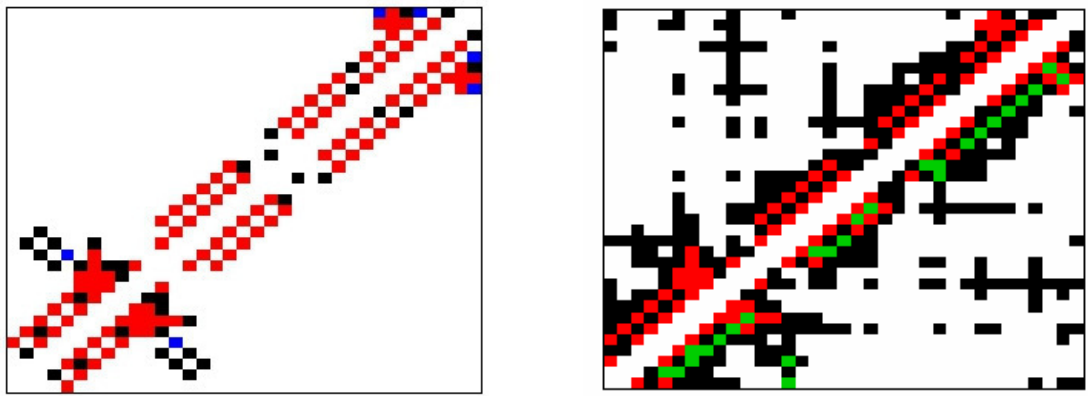

**Fig. 8 Left:** Contact map showing the 77 E edges. Right: Contact map of the native structure 2F4K. Color scheme: red=native shortcut, black=native edge, blue= neither native edge nor native shortcut, and green=hydrogen bonds.

## 4. Conclusion

The network of shortcut edges present in a native-state protein (SCN0) is proposed as an effective structural abstraction of protein molecules for folding purposes. Shortcut edges are identified by the EDS algorithm on a Protein Residue Network of a native-state protein (PRN0) [16, 24]. The EDS algorithm is based on an intuitive message passing algorithm on a social network [27, 28]. Yet, SCN0s are able to capture the essential structural characteristics of proteins (section 3.1), and to produce single-value descriptors of protein topology that correlate significantly with protein folding kinetics data irrespective of protein size (section 3.2). Notably, the logarithm form of SCN0 contact order (SCN0_lnCO) correlates significantly with protein kinetic rates regardless of size, and the clustering coefficient of SCN0 ($C_{SCN0}$) correlates significantly with folding rates, transition-state placement and stability of two-state folders.

The relevance of SCN0s to protein folding does not end with statistics, but with specifics. Within the Restricted Binary Collision (RBC) model, $C_{SCN0}$ could identify initial folding reaction points and suggest reasonable folding pathways for 12 small single-domain two-state model proteins (section 3.3) and three non-canonical proteins (section 3.4): ACBP which is sometimes classified as a non-two-state folder, Top7 which is a small and yet non-cooperative folder with a stable intermediate state, and DHFR which is 159 residues in length and has a dumb-bell domain structure. RBC starts with isolated pre-formed secondary structures which are incrementally coalesced to form the whole native structure. The coalescence rule maximizes the sum of the clustering coefficients of the shortcut networks ($C_{SCN0}$) within substructures of a partially formed native structure, whilst allowing for restricted interaction between substructures.

The bias showed by $C_{SCN0}$ in a native-state structure is detectable in non-native structures as well. This is demonstrated with the MD generated Villin headpiece structures. Further, a global examination of SCNs of non-native structures provides a novel view of the folding landscape. Such an analysis revealed three fold success factors for the Villin headpiece peptide, and confirmed a proposal made in [41].

These SCN0 results are edifying because they not only provide another line of evidence confirming the dominant influence of native-state topology on folding of globular proteins, but they do so in a way that is scalable. Such high-throughput analysis of folding pathways could reveal other folding patterns at the mesoscopic level. We have already detected a folding pattern that works well for the proteins tested; this folding pattern is revealed by RBC's preference to combine adjacent β-strands early (section 3.3).

There are a number of areas ripe for further study. When discussing transition-state placement (section 3.2), we mentioned how key folding residues could be distinguished by rigidity analysis using network characteristics of nodes [83]. Curiously, these key folding residues overlap only minimally with residues that have large positive φ values by the protein engineering method [46]. One reason for this is residues that have large positive φ values tend to be located in turns, and turn residues tend to have fewer network connections. The question how residues with large positive φ values work with the key folding residues to form the transition-state structures could be enlightened with further network analysis.

For a number of proteins, reasons were given to support the $C_{SCN0}$ folding pathways that diverged from the **E** folding pathways (section 3.3). However, the **E** folding pathways are not "optimized": a different, more reasonable, folding pathway could be generated with other energy functions, or by multiple samplings of Conformation energy and normalization.

The $C_{SCN0}$ folding pathway for 7DFR is sensitive to secondary structure delineation, and was found with an *ad hoc* modification to Eq. 1 (section 3.4). While this result demonstrates the potential of RBC for multi-domain proteins, further study and more tests are required to generalize this method for larger proteins.

## Acknowledgements

This work was made possible by the facilities of the Shared Hierarchical Academic Research Computing Network (SHARCNET:www.sharcnet.ca) and Compute/Calcul Canada. Thanks to C.N. Rowley for helpful discussions.

# Supporting Information

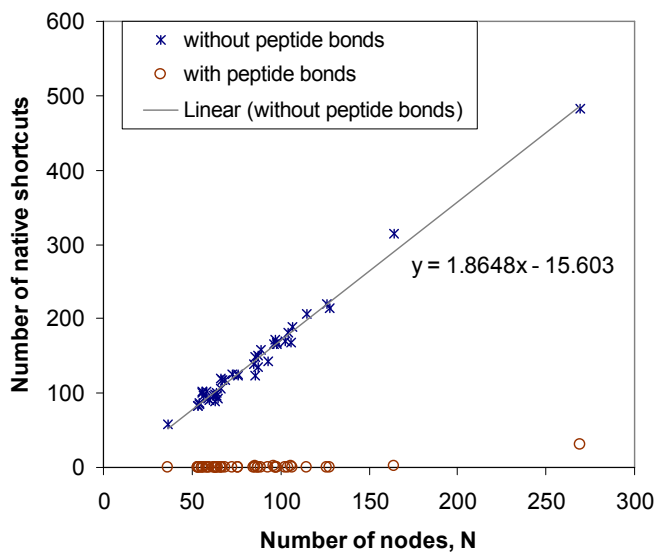

**Fig. S1** Number of shortcuts discovered by EDS on PRNs constructed with and without peptide bonds. When peptide bonds are excluded, the $|SC|\approx 2N$ relationship observed previously in [16] between the number of native shortcut edges $|SC|$ and the protein chain length $N$, is observed.

**Table S1** Description of proteins used in this study.

| PDB ID | Chain | Residue range | Secondary structure | | | Protein description |
|---|---|---|---|---|---|---|
| | | | H | S | T | |
| 1MBC | A | 1...153 | 121 | 0 | 32 | Carbon-monoxy (FE 11)-Myoglobin |
| 1BIN | A | 1...143 | 107 | 0 | 36 | Leghemoglobin A |
| 2ABD (Model 1) | A | 1...86 | 60 | 0 | 26 | Bovine acyl-coenzyme A binding protein |
| 1ST7 (Model 1) | A | 1...86 | 60 | 0 | 26 | Yeast acyl-coenzyme A binding protein |
| 1BDD | A | 1…60 | 36 | 0 | 24 | BdpA: Immunoglobulin-binding B domain of protein A |
| 2F4K | A | 42…76 | 25 | 0 | 10 | Chicken Villin subdomain HP-35, K65(NLE), N68H, K70(NLE) |
| 1COA | I | 20...83 | 13 | 21 | 30 | Chymotrypsin Inhibitor 2 |
| 2CI2 | I | 19...83 | 13 | 22 | 30 | Serine proteinase inhibitor (Chymotrypsin) CI-2 from barley seeds |
| 6PTI | A | 1...56 | 16 | 15 | 25 | Bovine pancreatic proteinase inhibitor (Trypsin) |
| 2IGD | A | 1...61 | 13 | 26 | 22 | Protein G Igg-Binding domain |
| 1GB1 (Model 1) | A | 1…56 | 13 | 26 | 17 | GB1: Immunoglobulin-binding domain of Streptococcal protein G |
| 1MHX | A | 1…65 | 16 | 28 | 21 | NuG1: Redesigned B1 domain of protein G |
| 1MI0 | A | 1…61 | 15 | 26 | 20 | NuG2: Redesigned B1 domain of protein G |
| 2PTL (Model 1) | A | 17…78 | 15 | 35 | 12 | LB1: Immunoglobulin light-chain binding domain of protein L |
| 2EZN (Model 1) | A | 1...101 | 14 | 58 | 29 | HIV-Inactivating protein Cyanovirin-N |
| 2CRT | A | 1...60 | 0 | 29 | 31 | Cardiotoxin III from Taiwan cobra |
| 1SHG | A | 6...62 | 3 | 26 | 28 | α-spectrin SH3 domain |
| 1SRM (Model 1) | A | 9...64 | 6 | 16 | 34 | src Tyrosine Kinase SH3 domain |
| 2KJV (Model 1) | A | 1...101 | 28 | 14 | 59 | Ribosomal protein S6 (Wild-type) |
| 2KJW (Model 1) | A | 1...96 | 30 | 23 | 43 | Ribosomal protein S6 (permutant P54-55) |
| 1QYS | A | 3...94 | 33 | 37 | 22 | Top7 de novo protein |
| 7DFR | A | 1...159 | 39 | 52 | 68 | E. coli Dihydrofolate reductase |

**Note S1 Shortcut identification by EDS**

Let $\mathcal{U}(p)$ be the union of the direct neighbors of all nodes in a path $p = \langle s\ldots a\ldots b\ldots c\rangle$, where $s$ is the source node, and $c$ is the most recent node appended to $p$. EDS adds a node $z$, which is not already in $p$, to $p$ if $z$ is closest (in Euclidean distance) to the target node amongst all nodes in $\mathcal{U}(p)$. Each time EDS adds a node not already in a path, the new node is given a *timestamp*. Let $T(v)$ be a positive integer denoting the timestamp of node $v$ on an EDS path. Node timestamps start at 1 for the source (first) node in a path, and increase by one for each *new* node added to a path. If $z$ is adjacent to $c$, then $z$ is appended to $p$, i.e. $p = \langle s\ldots a\ldots b\ldots c, z\rangle$. Otherwise, EDS backtracks to the node in $p$ with the largest timestamp that is adjacent to $z$. If for instance, $z$ is in the direct neighborhoods of both $a$ and $b$ (and $a \neq b$), EDS backtracks to $b$ to reach $z$ since $T(b) > T(a)$, and $p = \langle s\ldots a\ldots b\ldots c\ldots b, z\rangle$. Fig. S2 demonstrates the construction of three EDS paths.

An edge $(u, v)$ is a *shortcut* if and only if $T(v) = T(u) + 1$, and $v$ is adjacent to a node $w$ such that $T(w) < T(u)$. Shortcut edges help EDS avoid backtracking. For instance, without the shortcut (57, 54) in Fig. S2-left, EDS needs to backtrack to node 55, and then to node 40 to reach node 54. The condition $T(v) = T(u) + 1$ is necessary, so that the process of backtracking itself does not produce shortcut edges. For example, without this condition, (74, 77) in Fig. S2-center would be incorrectly identified as a shortcut edge for this path as a result of EDS backtracking from node 76 to reach node 77. Incidentally, (74, 77) is a shortcut in this PRN, but is identified as so by other EDS paths.

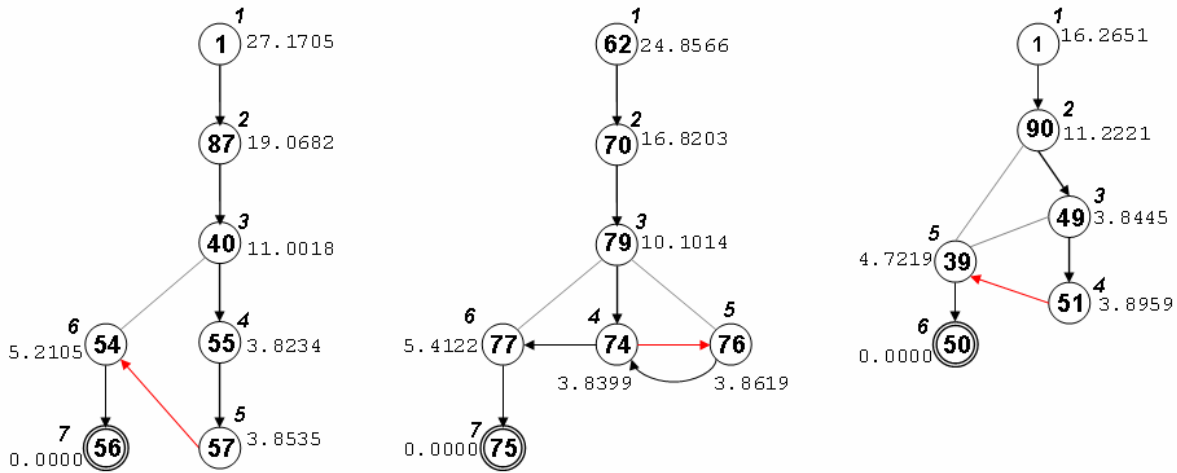

**Fig. S2** Three EDS paths from 2EZN PRNs. PRN edges are undirected, but the edges are oriented in the diagram in the direction they are traversed by EDS in the respective paths. Un-oriented edges are not traversed, but exist and play a role in determining whether an edge is a shortcut. Shortcut edges are marked in red. The real number beside each node is the node's Euclidean distance to the target node. The italicized integer beside each node $v$ is the node's timestamp, or $T(v)$ value. The leftmost EDS path is of length six, and is $\langle 1, 87, 40, 55, 57, 54, 56\rangle$. It has one shortcut edge: (57, 54). The center EDS path is of length seven, and is $\langle 62, 70, 79, 74, 76, 74, 77, 75\rangle$. It has one shortcut edge: (74, 76). The rightmost EDS path is of length five, and is $\langle 1, 90, 49, 51, 39, 50\rangle$. It has one shortcut edge: (51, 39).

**Table S2** The Zou-Ozkan dataset: 55 two-state proteins from ref. [32] by secondary structure content, and residue range where applicable.

| Structure | PIDs |
|---|---|
| α | 2PDD, 1BDD, 1L2Y, 1YCC, 1HRC, 1AYI, 1PRB, 1IDY, 1BA5, 2ABD, 1IMQ, 1VII, 2A3D, 1ENH, 256B, 1LMB-3(6…92) |
| α/β | 1POH, 1E65-A, 1APS, 1CIS, 1HZ6-A(2…64), 2CI2, 1COA, 1PBA, 1FKB, 1LOP, 2VIK, 1SPR-A, 1UBQ, 2ACY, 1URN, 1BF4, 1DIV_C, 2PTL, 1RIS, 1HDN, 1DIV_N |
| β | 1MJC, 2VKN, 1CSP, 1TEN-A(803...891), 1SHF-A, 1PIN, 1SRL, 1E0L, 1C8C, 1RLQ(9…64), 1FNF_9, 1WIT, 1QTU, 1K8M, 2AIT, 1NYF, 1SHG, 1PKS |

**Table S3** Secondary structure elements (SSEs) for the 16 proteins with examined folding pathways.

| PDB ID | SSEs listed from N- to C- termini for RBC | Common descriptors of the non-Turn SSEs |
|---|---|---|
| 1BDD | 0T 1H 2T 3H 4T 5H 6T | α1, α2, α3 |
| 2IGD | 0T 1S 2T 3S 4T 5H 6T 7S 8T 9S | β1, β2, α1, β3, β4 |
| 1GB1 | 0S 1T 2S 3T 4H 5T 6S 7T 8S | β1, β2, α1, β3, β4 |
| 1MHX | 0T 1S 2T 3S 4T 5H 6T 7S 8T 9S 10T | β1, β2, α1, β3, β4 |
| 1MI0 | 0T 1S 2T 3S 4T 5H 6T 7S 8T 9S 10T | β1, β2, α1, β3, β4 |
| 2PTL | 0S 1T 2S 3T 4H 5T 6S 7T 8S 9T | β1, β2, α1, β3, β4 |
| 2CI2 | 0T 1S 2T 3H 4T 5S 6T 7S 8T 9S | β1, α1, β2, β3, β4 |
| 1SHG | 0T 1S 2T 3S 4T 5S 6T 7S 8T 9S | β1, β2, β3, β4, β5 |
| 1SRM | 0T 1S 2T 3S 4T 5S 6T 7S 8T 9S | β1, β2, β3, β4, β5 |
| 2KJV | 0T 1S 2T 3H 4T 5S 6T 7S 8T 9H 10T 11S 12T | β1, α1, β2, β3, α2, β4 |
| 2KJW | 0T 1S 2T 3H 4T 5S 6T 7S 8T 9H 10T 11S 12T | β3, α2, β4, β1, α1, β2 |
| 2ABD | 0T 1H 2T 3H 4T 5H 6T 7H 8T | α1, α2, α3, α4 |
| 1ST7 | 0T 1H 2T 3H 4T 5H 6T 7H 8T | α1, α2, α3, α4 |
| 1QYS | 0T 1S 2T 3S 4T 5H 6T 7S 8T 9H 10T 11S 12T 13S 14T | β1, β2, α1, β3, α2, β4, β5 |
| 7DFR | 0S 1T 2H 3T 4S 5H 6T 7S 8T 9S 10T 11H 12T 13S 14H 15T 16S 17T 18S 19T 20S | β5, α1, β3, α2, β2, β1, α3, β4, α4, β6, β8, β7 [39] |
| 2F4K | 0T 1H 2T 3H 4T 5H 6T | α1, α2, α3 |

The SSE delineations in [13] (faculty.cs.tamu.edu/shsze/ssfold) are adopted for 1GB1, 1MHX, 1MI0, 2PTL and 2CI2. The SSEs are labeled such that *xSS* represents the $x^{th}$ SSE, with *x* starting at 0 for the first SSE at the N-terminus, and *SS* represents the type of SSE.

SSE delineations for the other proteins follow (residues belonging to a SSE are in parentheses):

**1BDD** : 0T(1…9), 1H(10…17), 2T(18…24), 3H(25…37), 4T(38…41), 5H(42…56), and 6T(57…60).
**2F4K**: 0T(42, 43), 1H(44…51), 2T(52…54), 3H(55…60), 4T(61...62), 5H(63…73), and 6T(74…76).
**2IGD**: 0T(1...5), 1S(6...13), 2T(14...17), 3S(18...24), 4T(25...27), 5H(28...40), 6T(41...46), 7S(47...51), 8T(52...55) and 9S(56...61) .
**1SHG**: 0T(6, 7), 1S(8...11), 2T(12...29), 3S( 30...35), 4T (36...40), 5S(41...46), 6T(47, 48), 7S(49...54), 8T(55...57), 9S(58...61), and 10T(62).
**1SRM**: 0T(9, 10), 1S(11, 12), 2T(13...32), 3S(33, 34), 4T(35...41), 5S(42...46), 6T(47...52), 7S(53...57), 8T(58...60), 9S(61, 62), and 10T(63, 64).
**2KJV**: 0T(1...3), 1S(4...7), 2T(8...15), 3H(16...32), 4T(33...39), 5S(40...42), 6T(43...60), 7S(61...63), 8T(64...70), 9H(71...81), 10T(82...88), 11S(89...92) and 12T(93...101).
**2KJW**: 0T(1...6), 1S(7...13), 2T(14, 15), 3H(16...27), 4T(28...31), 5S(32...39), 6T(40...45), 7S(46...51), 8T(52...57), 9H(58...75), 10T(76...82), 11S(83, 84) and 12T(85...96).
**2ABD** [71] and **1ST7** [74]: 0T(1, 2), 1H(3...15), 2T(16...20), 3H(21...36), 4T(37...51), 5H(52...62), 6T(63, 64), 7H(65...84) and 8T(85, 86).
**1QYS**: 0T(3), 1S(4...11), 2T(12...15), 3S(16...23), 4T(24...27), 5H(28...43), 6T(44...46), 7S(47...53), 8T(54...56), 9H(57...73), 10T(74...77), 11S(78...84), 12T(85,86), 13S(87...93) and 14T(94).

**7DFR** [39]: 0S(1...8), 1T(9...23), 2H(24...35), 3T(36...38), 4S(39...42), 5H(43...50), 6T(51...57), 7S(58...63), 8T(64...72), 9S(73...75), 10T(76), 11H(77...86), 12T(87...90), 13S(91...95), 14H(96...104), 15T(105...107), 16S(108...116), 17T(117...131), 18S(132...141), 19T(142...149), 20S(150...159).

**Note S2 Fitness of a Conformation X in RBC**

Eq.1 is compared with Eq. 2. The first term in Eq. 1 allows interactions between multi-SSE SSUs when calculating Conformation fitness. Eq. 2 ignores such interactions, and is the sum of each SSU evaluated in isolation.

$$\mathbf{F}(X) = f(RSU = \bigcup_{i=1}^{q} SSU_i) + \sum_{i=1}^{p} f(RS_i = SSU_i) \qquad \text{Eq. 1}$$

where *RSU* is the union of the residues belonging to the *q* multi-SSE SSUs in Conformation X, $RS_i$ is the set of residues belonging to $SSU_i$, and *f* is either the structural or energetic fitness function (section 2.7).

Note that the union of residues belonging to a set of multi-SSE SSUs (*RSU* in Eq. 1) need not be equal to the residues of the SSU containing all the non-Turn SSEs in the set of multi-SSE SSUs. In the case of 1SHG for example, the union of residues in SSU(1S 3S) and residues in SSU(5S 7S) excludes residues in SSE 4T, but the set of residues for SSU(1S 3S 5S 7S) does contain the residues in SSE 4T.

$$\mathbf{F}(X) = \sum_{i=1}^{n} f(SSU_i) \qquad \text{Eq. 2}$$

where $SSU_i$ is the $i^{th}$ SSU of Conformation X which has *n* SSUs, and *f* is either the structural or energetic fitness function.

The difference between Eq. 1 and Eq. 2 is demonstrated with Fig. S14, which depicts the folding pathway for 1SHG. Observe that $C_{SCN0}$ of a beta-strand SSE is 0.0. By Eq. 2, Conformation(1S 3S)(5S 7S)(9S) has $C_{SCN0}$ = 0.2381+0.4762 = 0.7143, and is fitter than Conformation(1S)(3S 5S 7S)(9S) which has $C_{SCN0}$=0.3216, and also fitter than all other accessible Conformations from Conformation(1S)(3S)(5S 7S)(9S). Thus, Eq. 2 produces a different folding pathway than Eq. 1 for 1SHG (Table S4). The folding pathway proposed by Eq. 2 for 1SHG is less reasonable since the central three-stranded β-sheet observed early in folding does not form until the final fold step (section 3.3).

Different $C_{SCN0}$ folding pathways are generated by Eq. 2 for 1SRM, 2KJW and 1QYS. Since the Eq. 2 pathways in Table S4 are less reasonable than those generated with Eq. 1, they argue for the use of Eq. 1 over Eq. 2.

**Table S4** Comparison of $C_{SCN0}$ folding pathways generated with Eq. 1 and Eq. 2

| PDB ID | Eq. 1 | Eq. 2 |
|---|---|---|
| 1SHG | (β1 ((β2 (β3 β4)$_1$ )$_2$ β5)$_3$ )$_4$ | ((β1 β2)$_2$ ((β3 β4)$_1$ β5)$_3$)$_4$ |
| 1SRM | (β1 (((β2 β3)$_1$ β4)$_2$ β5)$_3$ )$_4$ | ((β1 (β2 β3)$_1$)$_3$ (β4 β5)$_2$)$_4$ |
| 2KJW | ((((β3 α2)$_2$ (β4 β1)$_1$)$_3$ α1)$_4$ β2)$_5$ | (((β3 α2)$_2$ (β4 β1)$_1$)$_4$ (α1 β2)$_3$)$_5$ |
| 1QYS | ((β1 β2)$_1$ (α1 (β3 (α2 (β4 β5)$_2$ )$_3$ )$_4$ )$_5$ )$_6$ | ((β1 β2)$_1$ ((α1 β3)$_3$ (α2 (β4 β5)$_2$)$_4$)$_5$)$_6$ |

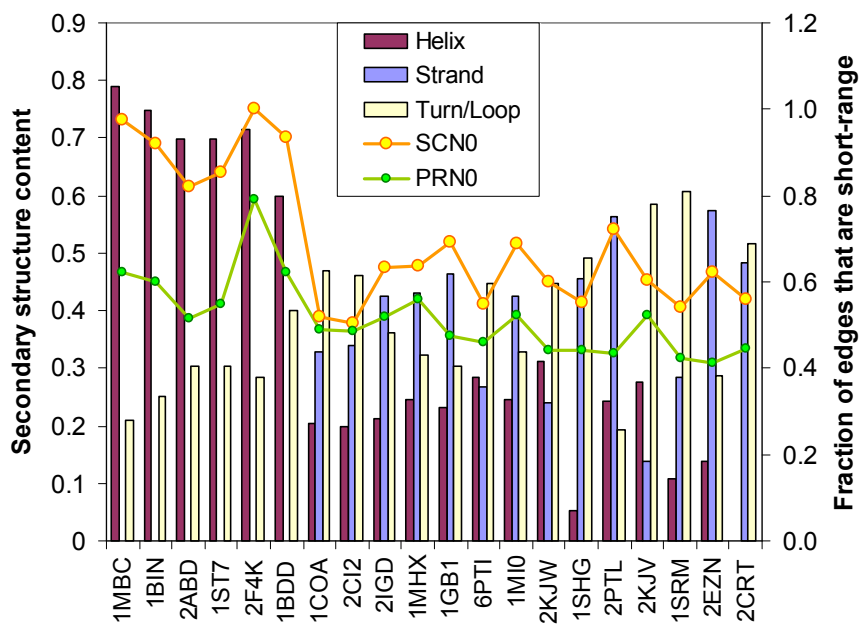

**Fig. S3** Secondary structure composition of 20 protein chains, and the fraction of SCN0 (PRN0) edges that are short-range (≤ 10 sequence positions apart). Almost all (>82%) of the native shortcuts in the six α-rich proteins are short-range, while an average of 60% of native shortcuts are short-range in the other proteins.

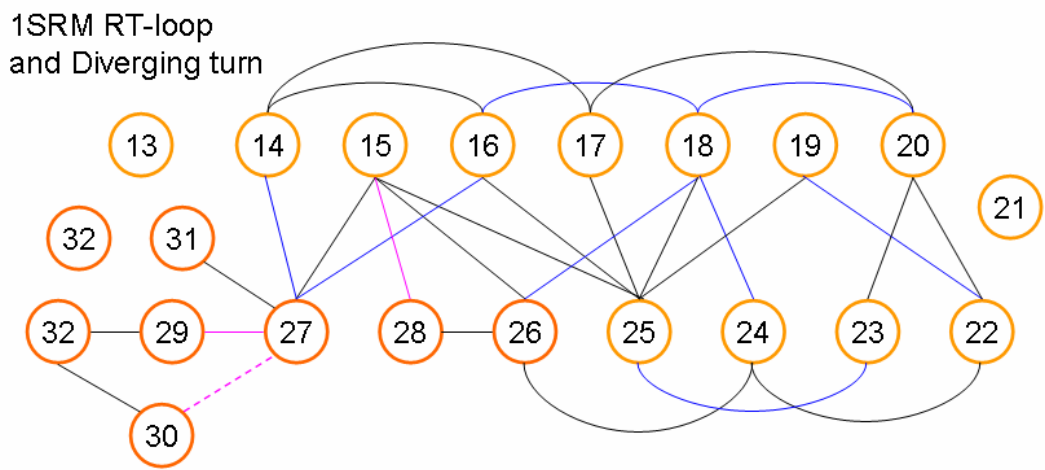

**Fig. S4** Illustration of shortcut edges and hydrogen bonds within the RT-loop and diverging turn of 1SRM. Circles represent PRN nodes (protein residues), black or blue solid lines represent shortcut edges, pink solid lines represent shortcut edges that are also hydrogen bonds, and pink dashed lines represent hydrogen bonds. Only the inter-molecular hydrogen bonds between non-adjacent residues reported by HBPlus [30] are shown. Circles are colored orange and dark orange to represent the RT-loop and diverging turn residues respectively. The diagram does not reflect the geometrical realities (lengths and angles) of the protein substructure.

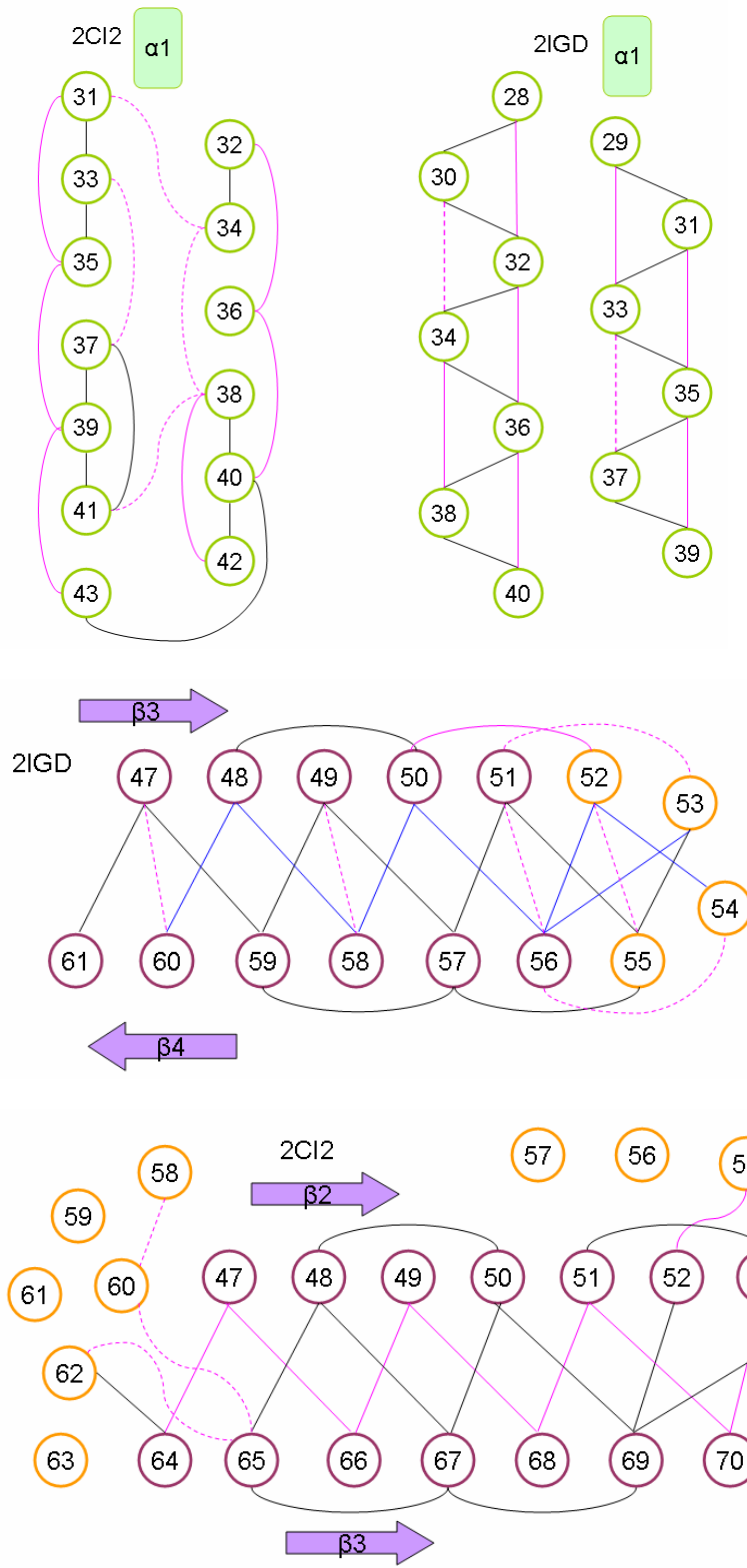

**Fig. S5** Illustration of shortcut edges and hydrogen bonds within two α-helix structures (top), an anti-parallel β-strand pair structure (middle), and a parallel β-strand pair structure (bottom). Circles represent PRN nodes (protein residues), black or blue solid lines represent shortcut edges, pink solid lines represent shortcut edges that are also hydrogen bonds, and pink dashed lines represent hydrogen bonds. Only the inter-molecular hydrogen bonds between non-adjacent residues reported by HBPlus [30] are shown. Circles are colored green, purple and orange for helix, strand and turn/loop residues respectively. The diagrams do not reflect the geometrical realities (lengths and angles) of the protein substructures.

Shortcut edges are shortest, sequence distance wise, within the α-helix structure, and longest within the parallel β-strand pairs. There is a gradual increase in sequence distance of shortcut edges within the (anti-parallel) β-hairpin from the turn end.

There is more overlap between shortcut edges and hydrogen bonds within the α-helix and parallel β-strand pair structures, than within the anti-parallel β-strand pairs. These peculiarities of shortcut edges and hydrogen bonds in the three secondary structure forms help to explain the positions of 1COA and 2CI2 (each of which has a parallel β-strand pair of some length) in Fig. S6, and the spike in *scse_hb* for both 2KJV and 2KJW (which have two α-helices each).

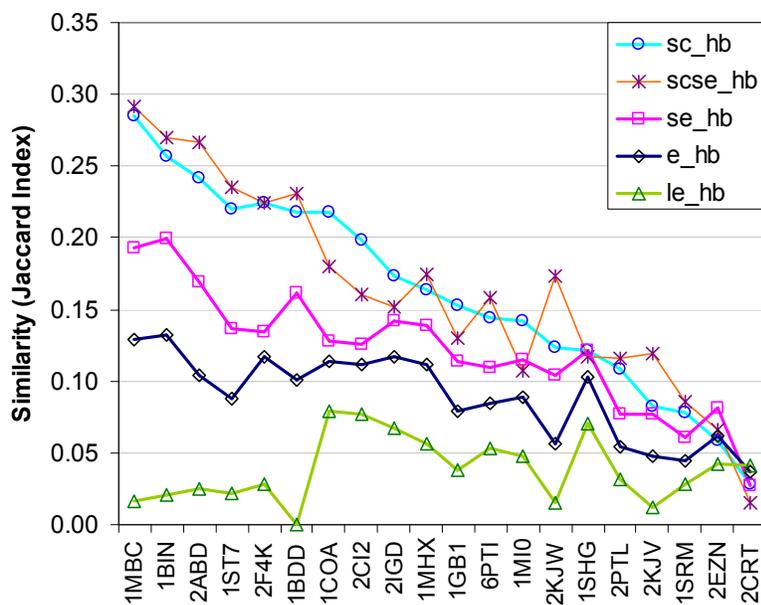

**Fig. S6** Similarity (measured by the Jaccard Index) between the set of hydrogen bonds (hb) obtained with HBPlus [30] on the (cleaned) PDB file of a protein, and five edge sets: all PRN0 edges (*e*), long-range PRN0 edges (*le*), short-range PRN0 edges (*se*), SCN0 edges (*sc*) and short-range SCN0 edges (*scse*).

For the 20 proteins examined, all inter-molecular hydrogen bonds between non-adjacent residues identified by HBPlus [30] for a protein are contained within the protein's PRN0 edge set. The shortcut edge sets are most similar to the hydrogen bond sets. The Jaccard Index computed with the shortcut set (*sc_hb*) is significantly (one-sided paired t-test p-value < 0.01) larger than the Jaccard Index computed with the set of short-range edges (*se_hb*), but not significantly different from the Jaccard Index computed with the set of short-range shortcut edges (*scse_hb*). Unlike shortcut edges, the hydrogen bonds of the proteins examined produce networks with negligible clustering coefficients.

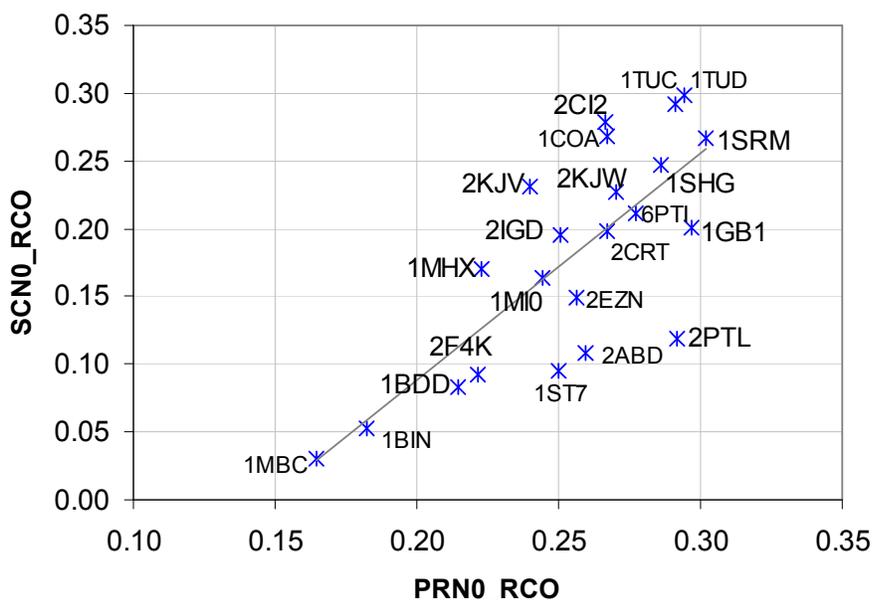

**Fig. S7** Relative Contact Order (RCO) calculated with native PRN and with native SCN edges only. The α-rich proteins (1MBC, 1BIN, 2ABD, 1ST7, 1BDD and 2F4K) are expected to fold faster than the β-rich proteins (1SRM, 1SHG and 2CRT). And indeed they have smaller RCO values, and occupy the lower left quadrant of the plot. The β-rich proteins tend to congregate in the upper right quadrant.

The SCN0_RCO and PRN0_RCO values are strongly and positively correlated (Pearson cor. = 0.7562). However, there are two pairs of proteins in the plot for which the inverse relationship between folding rate and RCO is more faithfully upheld by SCN0 than by PRN0.

2KJV and 2KJW is one such instance. 2KJW folds slightly faster than 2KJV [61]. In the plot, 2KJW is clearly to the right of 2KJV (2KJW has a larger PRN0_RCO than 2KJV), but is almost at the same level as 2KJV (2KJW has a slightly smaller SCN0_RCO than 2KJV).

The other instance involves 1SHG and its circular permutant 1TUD. The distal-loop, which is crucial to wild-type α-spectrin SH3 domain folding, is incised at N47-D48 in 1TUD. 1TUD is less stable (unfolds more quickly) and (re-)folds more slowly than wild-type α-spectrin SH3 domain [68]. In the plot, 1TUD is slightly to the right of but clearly above 1SHG. 1TUC, another α-spectrin SH3 circular permutant, has the RT-loop incised at S19-P20. While 1TUC is more unstable, it folds faster than the wild-type [68]. Both SCN0_RCO and PRN0_RCO fail to reflect this in the plot; 1TUD and 1TUC have almost identical SCN0_RCO and PRN0_RCO values.

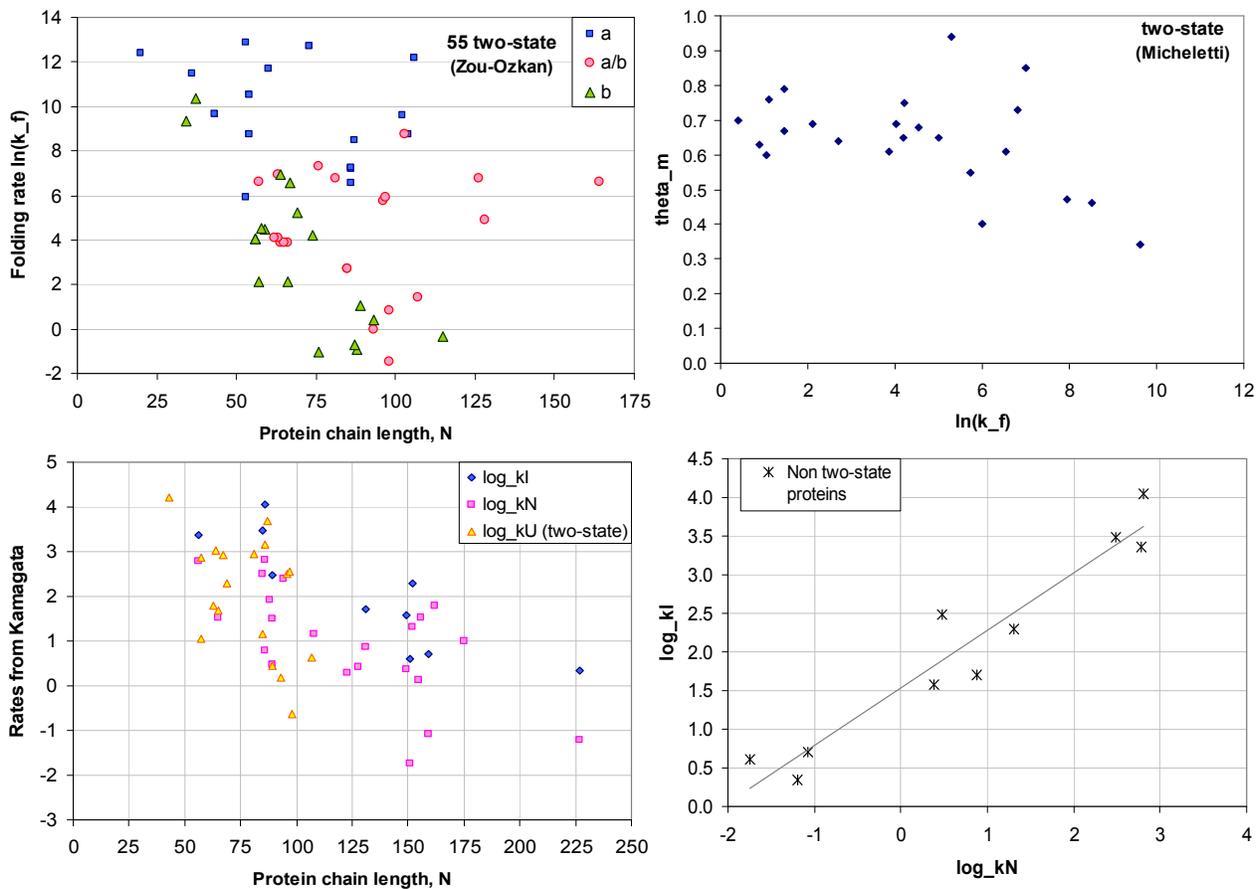

**Fig. S8** Top-left: Folding rate ln($k_f$), size and structure of 55 two-state proteins in the Zou-Ozkan dataset (Table S2). Top-right: The 23 two-state folders from the Micheletti dataset [14] exhibit a weak negative correlation (Pearson cor.= -0.4389, p-value= 0.0362) between transition-state placement (theta_m) and folding rate ln($k_f$). Bottom-left: Unfolding rates log($k_U$) of 19 two-state proteins, folding rates to native-state log($k_N$) of 23 non-two-state proteins, and folding rates to intermediate state log($k_I$) of 10 non-two-state proteins from the Kamagata dataset [33]. Bottom-right: These 10 non-two-state folders exhibit a strong positive correlation between log($k_I$) and log($k_N$).

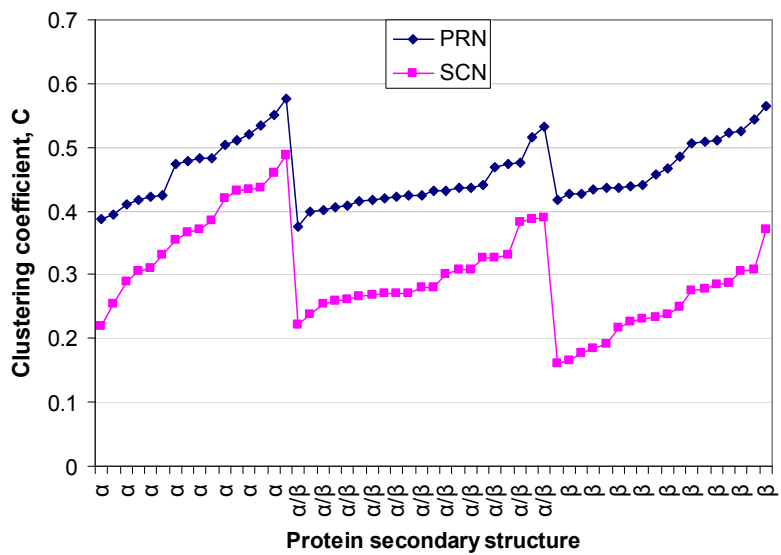

**Fig. S9** Network clustering coefficients calculated with PRN0 and with SCN0 edge sets for the 55 proteins in the Zou-Ozkan dataset, arranged by secondary structure content. $C_{SCN0}$ shows more sensitivity to differences in secondary structure content than $C_{PRN0}$.

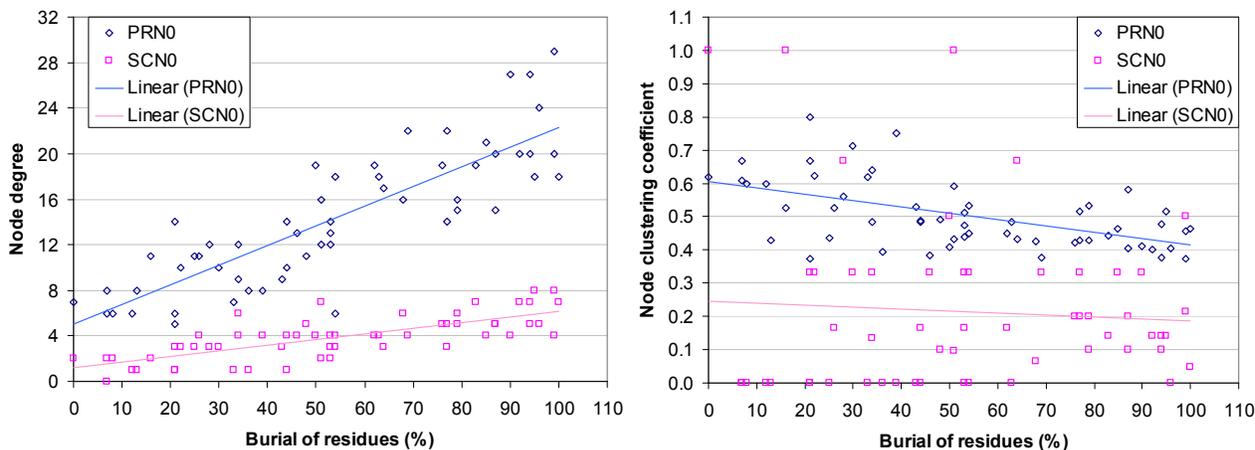

**Fig. S10** Burial of src SH3 domain residues in the native state (Burial data from Table 1 of ref. [64]) against 1SRM's PRN0 and SCN0 node statistics. Left: Pearson's correlation coefficient for PRN0 node degree and Burial of residues is 0.8448, and for SCN0 node degree and Burial of residues is 0.7442. Right: Pearson's correlation coefficient for PRN0 node clustering and Burial of residues is -0.5500, and for SCN0 node clustering and Burial of residues is -0.0667 (p-val=0.6249).

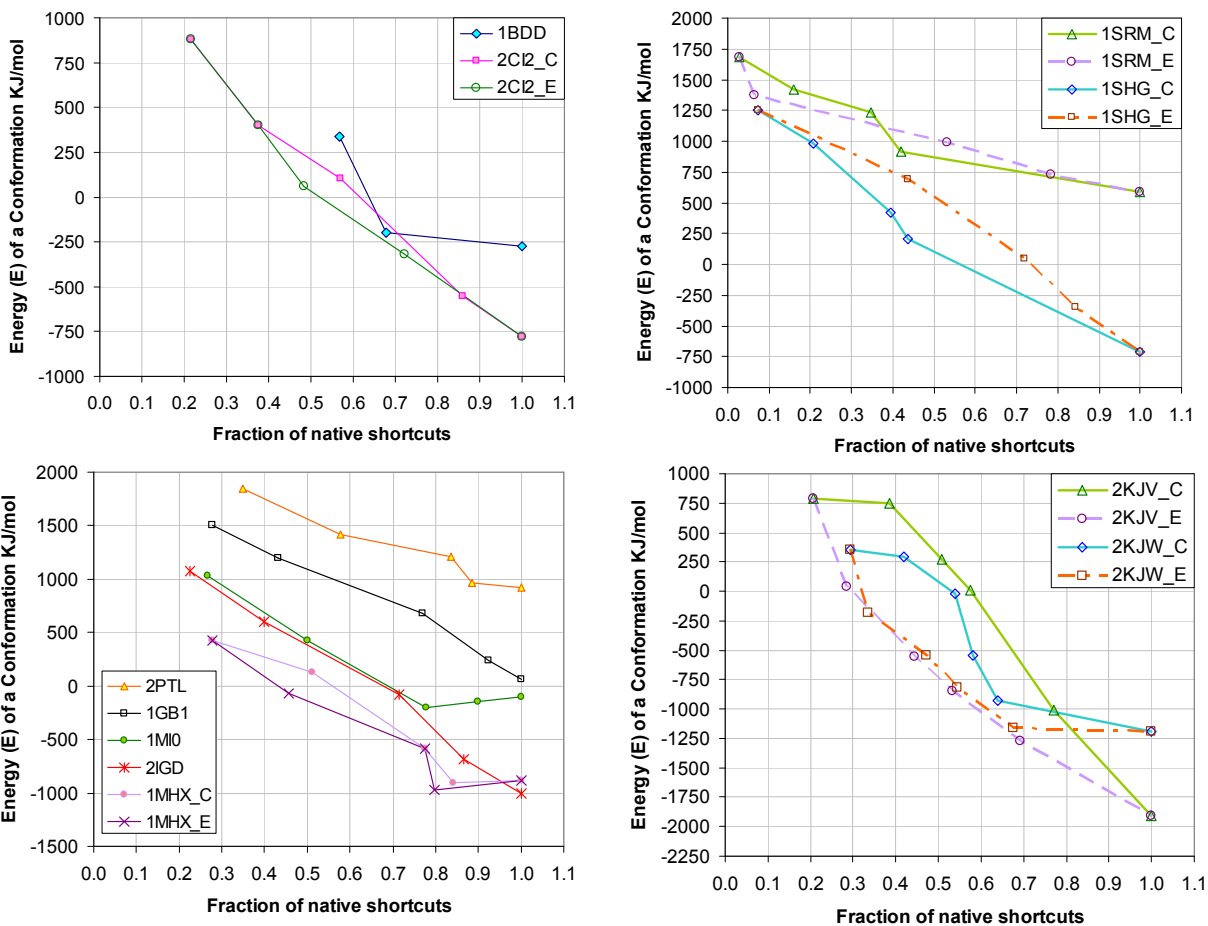

**Fig. S11** Energy of Conformations on $C_{SCN0}$ and on **E** identified folding pathways, plotted against their fraction of a protein's native shortcuts. When the plots of a protein differ, they are suffixed with _C and _E for $C_{SCN0}$ and **E** respectively. The $C_{SCN0}$ folding pathways exhibit a general downward energetic trend.

The following notes apply to **Figs. S12-S14**:

In the contact maps (adjacency matrices), shortcuts are marked by red cells, non-shortcut edges by black cells, and hydrogen bonds by green cells (in the lower triangle only).

In the search trees, $C_{SCN0}$ and **E** folding pathways are traced by green and blue arcs respectively. Grayed out boxes are infeasible Conformations. Dashed arcs connect accessible (within our RBC model) Conformations. To reduce clutter, not all accessible relations between Conformations are shown.

All *entry-point* Conformations (possible Conformations immediately prior to the native Conformation) are included in the diagram to depict the fitness landscape around the native Conformation.

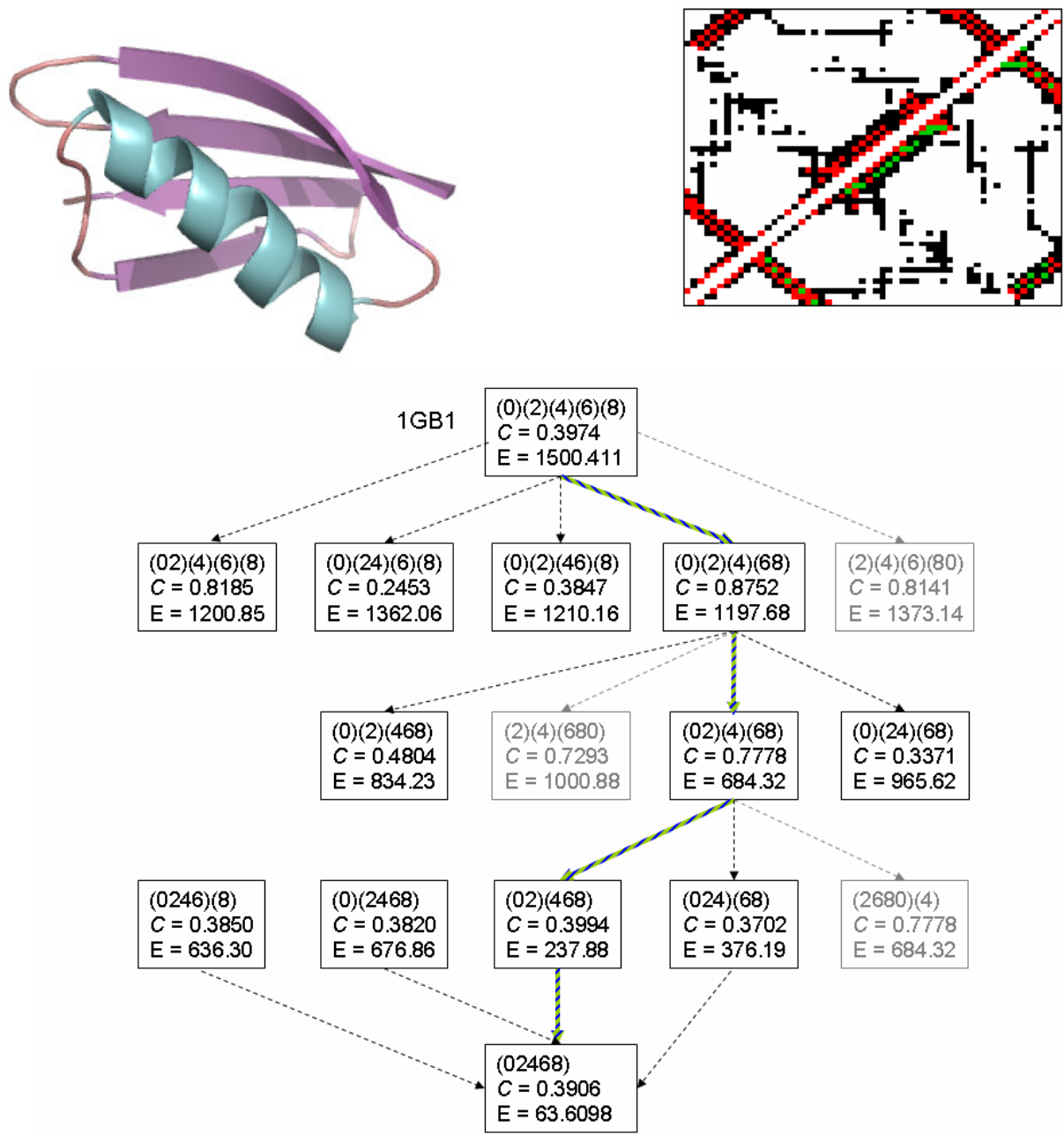

**Fig. S12** Top: Cartoon of 1GB1 and its PRN adjacency matrix. Bottom: 1GB1's $C_{SCN0}$ (green arcs), and **E** (blue arcs) folding pathways. The native Conformation has the lowest energy (E=63); both $C_{SCN0}$ and **E** folding pathways traverse the entry-point Conformation with the lowest energy (E=237).

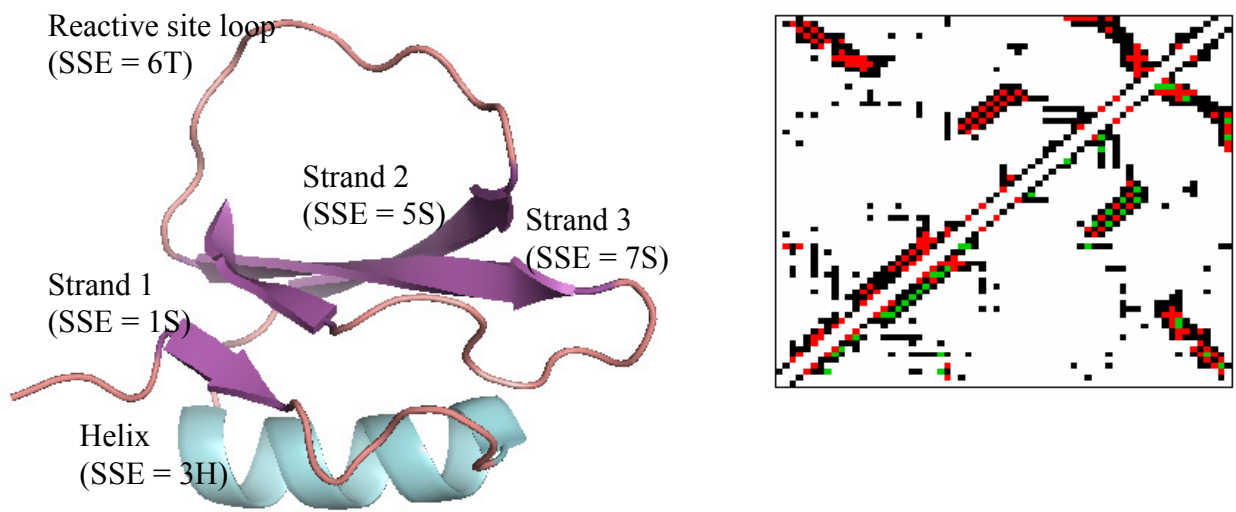
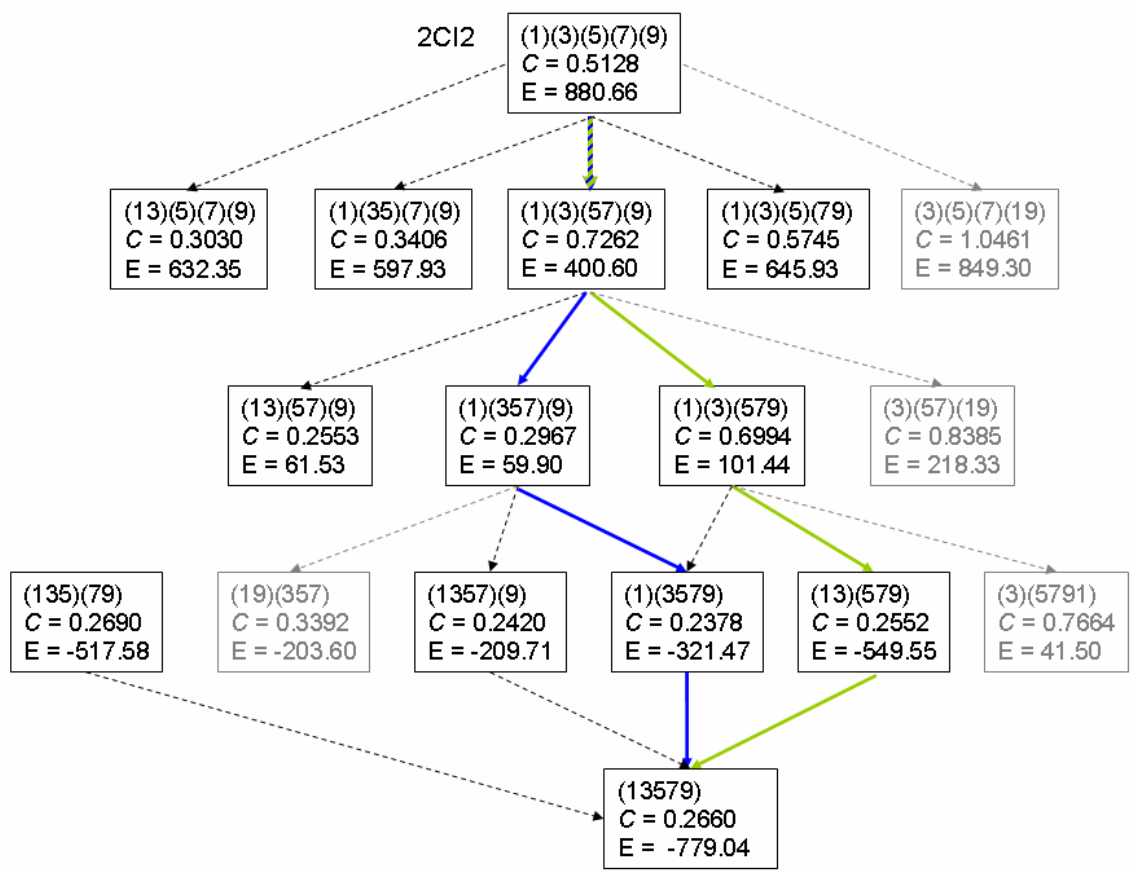

**Fig. S13** Top: Cartoon of 2CI2 and its PRN adjacency matrix. Bottom: 2CI2's $C_{SCN0}$ (green arcs), and **E** (blue arcs) folding pathways. The native Conformation has the lowest energy (E=-779); the $C_{SCN0}$ folding pathway traverses the entry-point Conformation with the lowest energy (E=-549).

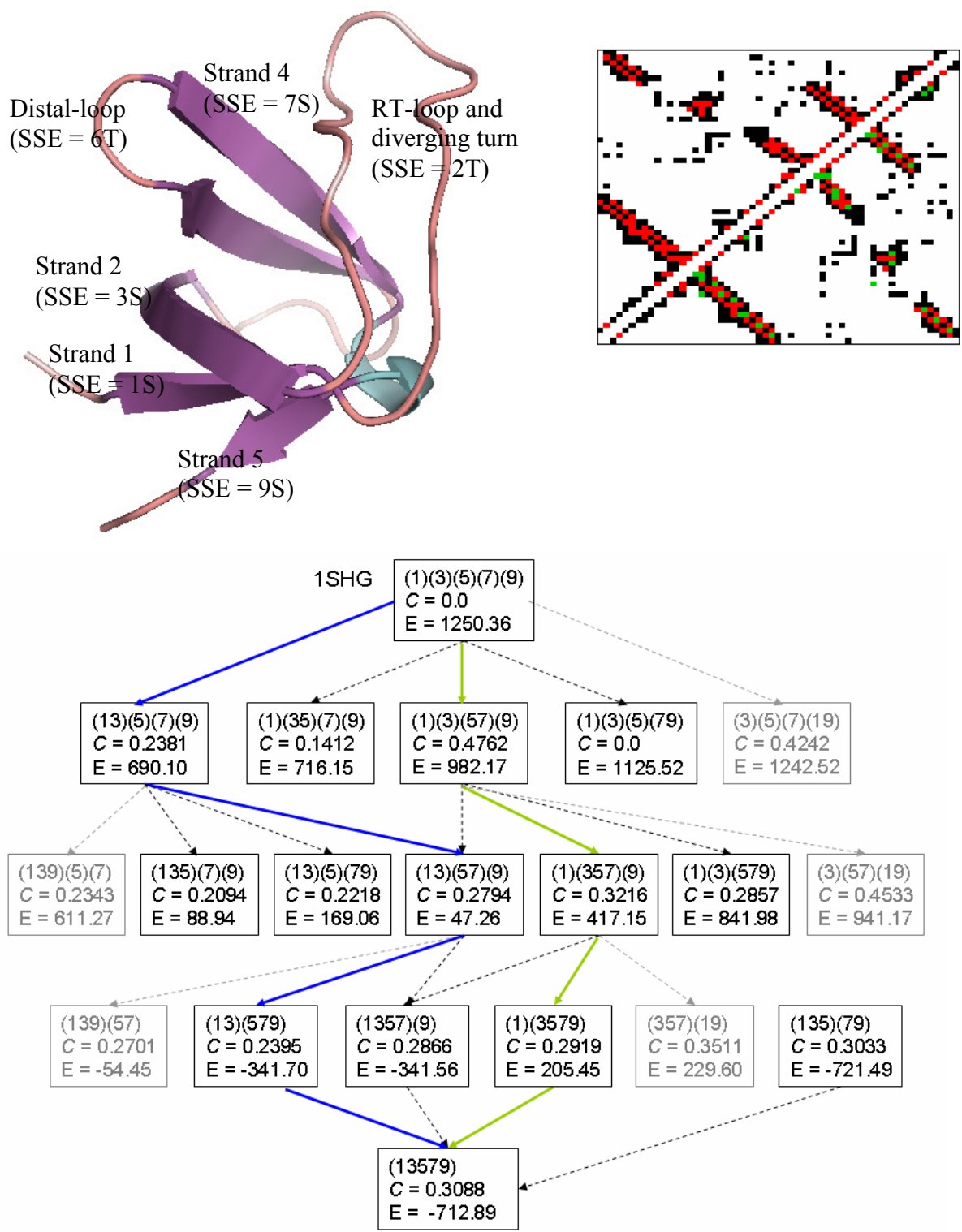

**Fig. S14** Top: Cartoon of 1SHG and adjacency matrix of 1SHG's PRN. Bottom: Construction of 1SHG's $C_{SCN0}$ ($C$) (green arcs) and energy (**E**) (blue arcs) folding pathways. Both the $C_{SCN0}$ and **E** folding pathways traverse entry-point Conformations with higher energy than the native Conformation which has the lowest energy in this case of 1SHG. The $C_{SCN0}$ folding pathway traverses the entry-point Conformation with the highest energy.

**Table S5** $C_{SCN0}$ folding pathway when N-C termini SSUs are permitted.

| PDB ID | $C_{SCN0}$ folding pathway with early N-C termini coupling allowed |
|--------|--------------------------------------------------------------------|
| 1BDD   | (1H (3H 5H)$_1$ )$_2$ |
| 2IGD   | ((3S (1S (7S 9S)$_1$ )$_2$ )$_3$ 5H)$_4$ |
| 1GB1   | (((0S 2S)$_2$ (6S 8S)$_1$ )$_3$ 4H)$_4$ |
| 2PTL   | (((0S 2S)$_1$ (6S 8S)$_2$ )$_3$ 4H)$_4$ |
| 1MHX   | (((1S 3S)$_1$ (7S 9S)$_2$ )$_3$ 5H)$_4$ |
| 1MI0   | (((1S 3S)$_1$ (7S 9S)$_2$ )$_3$ 5H)$_4$ |
| 2CI2   | (((9S 1S)$_1$ (5S 7S)$_2$ )$_3$ 3H)$_4$ |
| 1SHG   | (((9S 1S )$_2$ (5S 7S)$_1$ )$_3$ 3S)$_4$ |
| 1SRM   | (((9S 1S)$_1$ (3S 5S)$_2$)$_3$ 7S )$_4$ |
| 2KJV   | (((11S 1S)$_1$ (3H (5S 7S)$_2$)$_3$)$_4$ 9H)$_5$ |
| 2KJW   | ((((11S 1S)$_2$ 3H)$_3$ (5S 7S)$_1$)$_4$ 9H)$_5$ |
| 1QYS   | (((((1S 3S)$_1$ (11S 13S)$_2$)$_3$ 9H)$_4$ 7S)$_5$ 5H)$_6$ |

**Note S3**

Molecular Dynamics (MD) data for the Villin Headpiece subdomain is publicly available from https://simtk.org/home/foldvillin. The folding MD simulations (300 K) start from nine different 2F4K denatured structures: 0...8. Multiple trajectories are produced per initial configuration. Ensign *et al* [41] analyzed the trajectories by six criteria (related to the formation of the three helices and the pair-wise contact distances between F47, F51 and F58), and partitioned them into three sets. In the *unsuccessful* set are runs starting from initial structures 0, 1, 2, 3, 5, and 6. The trajectories generated with these starting structures either only briefly visited a folded structure or did not fold at all. In the *fast-successful* set are trajectories initiating with initial structures 4 and 7, and in the *slow-successful* set are trajectories starting with initial structure 8. The starting structures of the successful runs either folded much faster or folded to a significant extent. We chose trajectories from each initial structure at random, with a preference for longer runs. The runs in our sample (Table S6) have about the same number of snapshots each. There is no significant (t.test p-value 0.7378) difference in the average number of snapshots between the unsuccessful and successful runs.

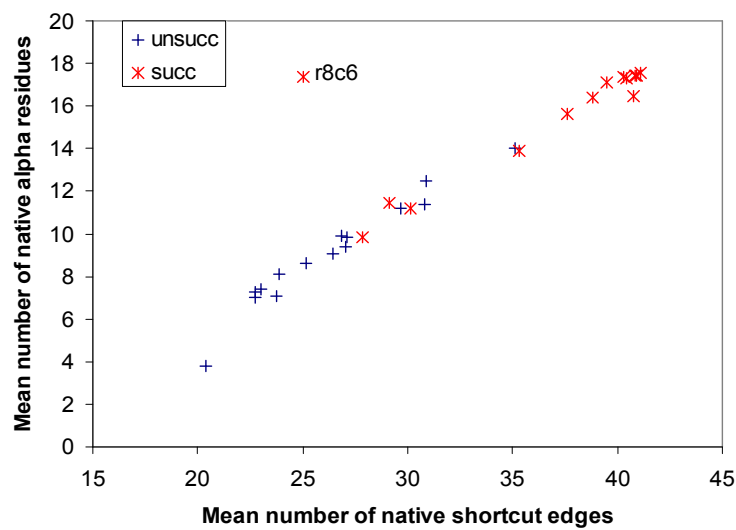

**Fig. S15** Pearson correlation between the average number of native shortcuts (SCN0 edges) and the average number of native alpha residues per run is 0.9131 (p-value 1.978E-12). Native alpha residues are those within the following residue ranges: E45...F51, S56...N60 and W64...K73 [41]. The successful runs (those starting with initial structures 4, 7 or 8) gravitate towards the top-right corner.

**Table S6** Number of snapshots in our sample of 2F4K MD runs. Runs are labeled as r*x*c*y* which means *y* trajectory with *x* initial structure. Total shortcuts is the number of unique shortcuts from all snapshots in a run.

| Unsuccessful | | | Successful | | |
| --- | --- | --- | --- | --- | --- |
| Run | Number of snapshots | Total shortcuts | Run | Number of snapshots | Total shortcuts |
| r0c1 | 37318 | 557 | r4c0 | 40117 | 283 |
| r0c2 | 39726 | 551 | r4c3 | 40122 | 303 |
| r0c3 | 40117 | 536 | r4c5 | 38131 | 527 |
| r1c0 | 38512 | 544 | r4c8 | 36936 | 261 |
| r1c1 | 37762 | 471 | r4c11 | 39717 | 271 |
| r2c2 | 38945 | 550 | r7c0 | 40133 | 225 |
| r2c4 | 36919 | 448 | r7c1 | 36927 | 426 |
| r2c23 | 38569 | 560 | r7c2 | 40146 | 227 |
| r3c2 | 37305 | 502 | r7c3 | 40135 | 217 |
| r3c7 | 36517 | 540 | r7c4 | 38119 | 437 |
| r3c8 | 39313 | 544 | r8c2 | 40123 | 330 |
| r5c0 | 39325 | 551 | r8c6 | 37317 | 545 |
| r5c18 | 39727 | 558 | r8c8 | 37725 | 368 |
| r6c1 | 40130 | 499 | r8c12 | 36915 | 271 |
| r6c2 | 38922 | 391 | r8c13 | 38915 | 371 |
| Mean ± std. dev. | 38607.13 ± 1184.19 | 520.13 ± 49.30 | Mean ± std. dev. | 38765.20 ± 1369.08 | 337.47 ± 105.91 |

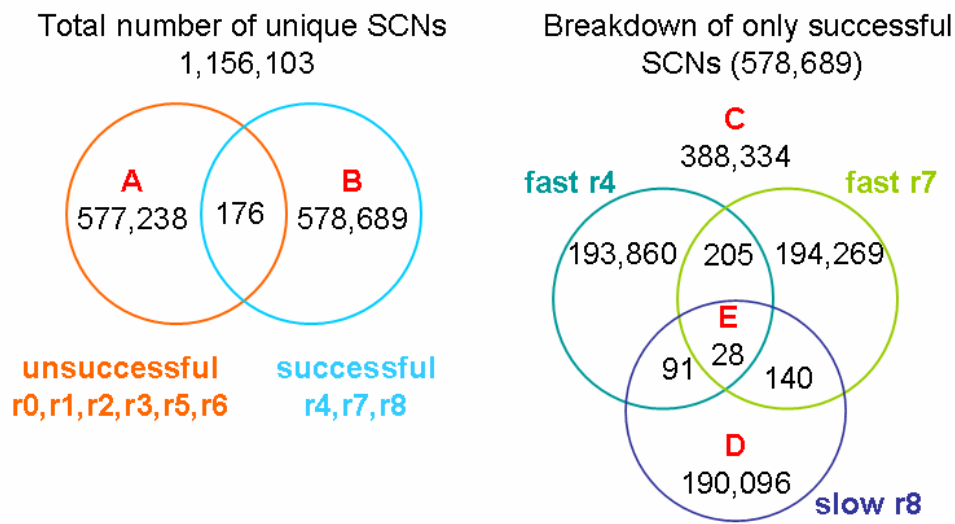

**Fig. S16** Number of SCNs in each of the five categories: (A) only unsuccessful (577,238), (B) only successful (578,689), (C) only fast (193,860+205+194,269=388,334), (D) only slow (190,096) and (E) all successful (28). There is minimal overlap between SCNs starting from different initial configurations. Out of the global set of 1,156,103 SCNs, only 645 were generated from more than one initial configuration. This testifies to the heterogeneity of the MD trajectories, at least where SCNs are concerned. The drastically smaller size of the E category reflects the funnel view of protein folding: that while there may be many possible folding pathways, they all need to pass through a narrow gate to reach the native-state.

**Table S7A** Aggregate statistics for SCNs in the five categories. The values reported are: mean ± std. dev. (above), and median (below).

| Metric | A | B | C | D | E |
|---|---|---|---|---|---|
| 1) $C_{SCN}$ | 0.30 ± 0.10<br>0.29 | 0.41 ± 0.10<br>0.43 | 0.42 ± 0.09<br>0.44 | 0.37 ± 0.11<br>0.39 | 0.47 ± 0.04<br>0.46 |
| 2) $C_{SCN}$ (1H 3H)(5H) - $C_{SCN}$ (1H)(3H 5H) | 0.17 ± 0.23<br>0.19 | 0.30 ± 0.23<br>0.31 | 0.31 ± 0.24<br>0.33 | 0.27 ± 0.22<br>0.28 | 0.40 ± 0.21<br>0.52 |
| 3) $C_{SCN}$ (1H 3H)(5H) > $C_{SCN}$ (1H)(3H 5H) | 77.08% | 88.46% | 88.84% | 87.66% | 100% |
| 4) \|SCLE\| | 6.19 ± 5.24<br>5.0 | 3.31 ± 3.79<br>2.0 | 2.84 ± 3.48<br>2.0 | 4.28 ± 4.20<br>2.0 | 1.36 ± 0.87<br>2.0 |
| 5) Proportion of SCLE that connect 1H with 3H | 10.87% | 17.72% | 18.83% | 16.19% | 45.81% |

Notes:
3) Proportion of SCNs in a set where $C_{SCN}$ (1H 3H)(5H) > $C_{SCN}$ (1H)(3H 5H) is true.
4) Number of long-range (>10 sequence positions) shortcut edges.
5) To capture the frequency of SCLE occurrence, this proportion is calculated over all SCLE appearing in a set of SCNs, not the unique SCLE in a set of SCNs. The E SCNs have in total 155 SCLE, but only two unique SCLE.

**Table S7B** P-values for the alternative hypotheses (Alt. H.) tested with R's t.test and wilcox.test commands.

| Metric | Alt. H. | p-value | Alt. H. | p-value | Alt. H. | p-value | Alt. H.* | p-value |
|---|---|---|---|---|---|---|---|---|
| 1) $C_{SCN}$ | A < B | 0<br>0 | C > D | 0<br>0 | A < D | 0<br>0 | C < E | 0<br>0.0020 |
| 2) $C_{SCN}$ (1H 3H)(5H) - $C_{SCN}$ (1H)(3H 5H) | A < B | 0<br>0 | C > D | 0<br>0 | A < D | 0<br>0 | C < E | 0.0158<br>0.0116 |
| 4) \|SCLE\| | A > B | 0<br>0 | C < D | 0<br>0 | A > D | 0<br>0 | C > E | 0<br>0.0319 |

*Note the vastly different sample sizes in these tests.

**Table S8** $C_{SCN0}$ and $T_{SCN0}$ values for 2F4K's SSUs. $T_{SCN0}$ is the number of native triangles in a structure; a native triangle is a cycle made by three distinct native shortcuts. Only initial structures for runs 4 and runs 7 have native triangles in 1H. The initial structure of a run is its first MD snapshot. Grayed out rows highlight initial structures that lead to successful runs.

| Structure | $C_{SCN0}$ ($T_{SCN0}$) | | | | | |
|---|---|---|---|---|---|---|
| | 1H | 3H | 5H | SSU(1H 3H) | SSU(3H 5H) | SSU(1H 3H 5H) |
| Initial for runs 0 | 0 | 0 | 0 | 0 | 0 | 0 (0) |
| Initial for runs 1 | 0 | 0.5 (1) | 0 | 0.1373 (1) | 0.1228 (1) | 0.0476 (1) |
| Initial for runs 2 | 0 | 0 | 0.6212 (4) | 0 | 0.3246 (4) | 0.1610 (4) |
| Initial for runs 3 | 0 | 0 | 0 | 0 | 0 | 0 (0) |
| Initial for runs 4 | **0.2917 (1)** | 0 | 0.2121 (1) | 0.2078 (2) | 0.0789 (1) | 0.1438 (3) |
| Initial for runs 5 | 0 | 0.5 (1) | 0 | 0.1373 (1) | 0.1754 (2) | 0.0762 (2) |
| Initial for runs 6 | 0 | 0 | 0.2727 (1) | 0 | 0.1579 (1) | 0.0857 (1) |
| Initial for runs 7 | **0.4167 (2)** | 0 | 0.2727 (1) | 0.1686 (2) | 0.1579 (1) | 0.1676 (3) |
| Initial for runs 8 | 0 | 0 | 0.3030 (2) | 0 | 0.1754 (2) | 0.0952 (2) |
| Native 2F4K | 0.7083 (4) | 1 (2) | 0.7727 (6) | 0.6358 (10) | 0.6579 (9) | 0.5184 (21) |

**Table S9A** The mean ± std. dev. (above) and median (below) development (r$T_{SCN0}$) of the three α-helices.

| Metric | A | B | C | D | E |
|---|---|---|---|---|---|
| 1) r$T_{SCN0}$ (1H) | 0.13 ± 0.22<br>0 | 0.21 ± 0.27<br>0 | 0.22 ± 0.28<br>0 | 0.20 ± 0.26<br>0 | 0.24 ± 0.39<br>0 |
| 2) r$T_{SCN0}$ (3H) | 0.31 ± 0.40<br>0 | 0.74 ± 0.38<br>1 | 0.80 ± 0.34<br>1 | 0.63 ± 0.42<br>1 | 1 ± 0<br>1 |
| 3) r$T_{SCN0}$ (5H) | 0.42 ± 0.32<br>0.33 | 0.64 ± 0.30<br>0.67 | 0.68 ± 0.28<br>0.67 | 0.56 ± 0.32<br>0.67 | 0.94 ± 0.10<br>1 |

**Table S9B** P-values for the alternative hypotheses tested with R's paired t.test and paired wilcox.test commands.

| Alternative Hypothesis | A | B | C | D | E |
|---|---|---|---|---|---|
| $rT_{SCN0}$ (1H) < $rT_{SCN0}$ (3H) | 0<br>0 | 0<br>0 | 0<br>0 | 0<br>0 | 0<br>0 |
| $rT_{SCN0}$ (1H) < $rT_{SCN0}$ (5H) | 0<br>0 | 0<br>0 | 0<br>0 | 0<br>0 | 0<br>0 |
| $rT_{SCN0}$ (3H) < $rT_{SCN0}$ (5H) | 0<br>0 | - | - | - | - |
| $rT_{SCN0}$ (3H) > $rT_{SCN0}$ (5H) | - | 0<br>0 | 0<br>0 | 0<br>0 | 0.0026<br>0.0013 |

**Table S9C** Significant (p-value < 0.01) Pearson (above) and Spearman (bottom) correlation coefficients between $rT_{SCN0}$(1H) and $rT_{SCN0}$ of all possible SSUs for 2F4K, computed over unique SCNs from *all* runs. $rT_{SCN0}$(1H) is most strongly and positively correlated with $rT_{SCN0}$(1H 3H).

| SSU | 3H | 5H | 1H 3H | 3H 5H | 1H 5H |
|---|---|---|---|---|---|
| Pearson cor. | 0.0999 | 0.0606 | 0.6240 | 0.0871 | 0.3891 |
| Spearman cor. | 0.0987 | 0.0581 | 0.5688 | 0.0851 | 0.3514 |

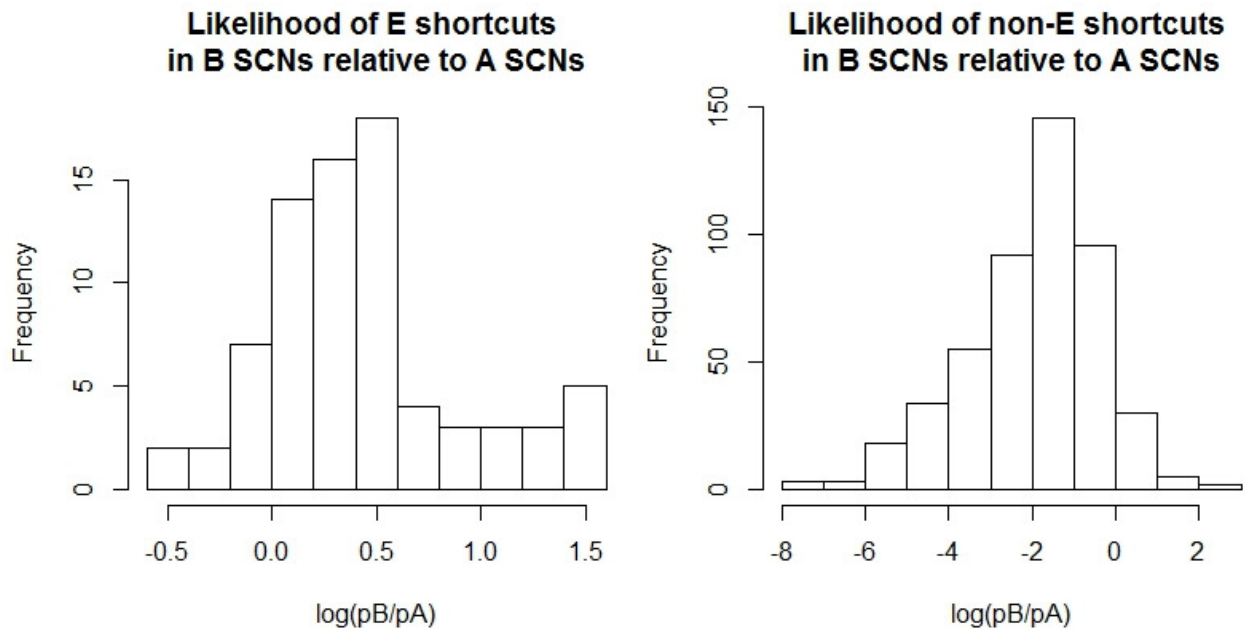

**Fig. S17** The majority of E edges lie above 0.0 (**left**), while the majority of non-E edges lie below 0.0 (**right**). pB = number of times an edge occurs in B /number of SCNs in B. pA= number of times an edge occurs in A /number of SCNs in A. The point of indifference at 0.0 occurs when pB=pA.